\documentclass[twocolumn,showpacs,preprintnumbers,amsmath,amssymb,prc]{revtex4-1}
\usepackage{graphicx}% Include figure files
\graphicspath{{./figures/}} %folder containing figures
\usepackage{epstopdf}
\usepackage{dcolumn}% Align table columns on decimal point
\usepackage{bm}% bold math
\usepackage{fp}

\begin{document}
	
	\preprint{ }
	
	\title{The  $^{12}$C(n, 2n)$^{11}$C cross section from threshold to 26.5 MeV}%
	%Force line breaks with \\
	
	\author{M. Yuly }
	%\altaffiliation[Also at ]{Physics Department, XYZ University.}%Lines break
	%automatically or can be forced with \\
	\email{mark.yuly@houghton.edu}
	%\homepage{http://www.houghton.edu/campus/facultystaff-directory/mark-yuly/}
	
	\author{ T. Eckert}%
	\author{ G. Hartshaw}%
	\affiliation{%
		Department of Physics, Houghton College, Houghton, New York 14744, USA
		%\\ This line break forced with \textbackslash\textbackslash
	}%
	
	\author{S. J. Padalino}
	\author{ D. N. Polsin}%
	\author{ M. Russ}%
	\author{ A. T. Simone }%
	\affiliation{
		Department of Physics, State University of New York, Geneseo, New York 14454,
		USA\\
	}%
	
	\author{C. R. Brune  }
	\author{T. N. Massey}%
	\author{C. E. Parker}
	\affiliation{
		Edwards Accelerator Laboratory, Department of Physics and Astronomy, Ohio
		University, Athens, Ohio 45701, USA \\
	}%
	
	\author{R. Fitzgerald  }
	\affiliation{
		National Institute of Standards and Technology, 100 Bureau Drive, Stop 8462,
		Gaithersburg, Maryland 20899-8462, USA \\
	}%
	
	\author{T. C. Sangster  }
	\author{S. P. Regan}%
	\affiliation{
		Laboratory for Laser Energetics, University of Rochester, Rochester, New York
		14623, USA \\
	}%
	\date{\today}% It is always \today, today,
	%  but any date may be explicitly specified

	\begin{abstract}
		The  $^{12}$C(n, 2n)$^{11}$C cross section was measured from just below
		threshold to 26.5 MeV using the Pelletron accelerator at Ohio
		University. Monoenergetic neutrons, produced via the $^3$H(d,n)$^4$He reaction,
		were allowed to strike targets of polyethylene and graphite. Activation of both
		targets was measured by counting positron annihilations resulting from the
		$\beta^+$ decay of $^{11}$C. Annihilation gamma rays were detected, both in
		coincidence and singly, using back-to-back NaI detectors.  The incident neutron
		flux was determined indirectly via $^{1}$H(n,p) protons elastically scattered from the
		polyethylene target. Previous measurements fall into  upper and lower
		bands; the results of the present measurement are consistent with the upper band.   
	\end{abstract}
	
	% Nucleon induced reactions +  Maybe  24.60.Dr Statistical
	%compound-nucleus reactions?
	% PACS, the Physics and Astronomy
	% Classification Scheme.
	%\keywords{Suggested keywords}%Use showkeys class option if keyword
	%display desired
	\maketitle
	
	\section{Introduction}
	
	The  $^{12}$C(n, 2n)$^{11}$C reaction may be a useful  and robust neutron diagnostic for measuring the areal density $\rho R$ of a deuterium-tritium (DT) implosion, which  is an important parameter in determining the implosion compression of an inertial confinement fusion (ICF) burn. Not only is the reaction sensitive to $\rho R$, but because the reaction is only sensitive to neutrons above 20 MeV, it is immune to primary neutrons (14.1 MeV) and down-scattered neutrons. To use this method, ultra-pure graphite disks placed within the ICF reaction chamber become activated by tertiary neutrons via the $^{12}$C(n, 2n)$^{11}$C reaction. The 511 keV gamma rays emitted by the $^{11}$C disk during positron annihilation are subsequently counted in an area far away from the target chamber and used to obtain the tertiary neutron yield. The diagnostic is well suited for the harsh EMP environment produced during an ICF implosion and high gamma and x-ray background  \cite{welch1988,glebov2003}. Furthermore, the $^{11}$C half-life is sufficiently long compared to the graphite extraction time which is on the order of a few minutes. This allows for the counting process to begin soon after the ICF shot prior to radioactive cooling of the graphite. It is important to note that the tertiary yield cannot be determined without a good knowledge of the $^{12}$C(n, 2n)$^{11}$C reaction cross sections in this energy range. Hence the importance of these new cross section measurements.

	%The  $^{12}$C(n, 2n)$^{11}$C reaction, with a threshold of $20.2957 \pm 0.0010$ MeV \cite{NuBase}, may be a
	%useful diagnostic tool for measuring the flux of tertiary neutrons produced by
	%inertial confinement fusion, since it is insensitive to the large number of 14
	%MeV primary neutrons produced in the primary DT fusion reaction.  Graphite disks
	%placed within the ICF reaction chamber would be activated by the tertiary
	%neutrons, which range up to about 30 MeV,  and yet survive even in the harsh
	%environment once ignition is achieved  \cite{welch1988,glebov2003}.
	
	A measurement of the  $^{12}$C(n, 2n)$^{11}$C reaction in this energy range is
	also important for calculations of the rate of cosmogenic $^{11}$C production,
	since the uncertainty in this cross section represents the largest source of
	systematic error in these calculations.  In turn, the presence of cosmogenic
	$^{11}$C in deep underground mines limits the detectability of \textit{pep} and
	CNO solar neutrinos in several neutrino experiments \cite{galbiati2005}.
	
	Figure 1 shows the previous measurements in the energy range between threshold (at $20.2957 \pm 0.0010$ MeV \cite{NuBase})
	and 35 MeV, as well as predicted cross sections from Dimbylow
	\cite{dimbylow1980}, which are from a nuclear optical model calculation using
	fits to experimentally measured total, elastic and inelastic cross sections. 
	The cross sections tend to follow two separate bands which differ by as much as
	a factor of two across the neutron energy range of  interest. The upper band
	comprises measurements from Anders et al. \cite{anders1981} and Welch et al.
	\cite{welch1981} and calculations of Dimbylow \cite{dimbylow1980}, while the
	lower band is  the measurements from Brolley et al. \cite{brolley1952}, Brill et
	al. \cite{brill1961}, Soewarsono et al \cite{soewarsono1992}, and Uno et al.
	\cite{uno1996}.
	
	\begin{figure}
		\includegraphics[width=\linewidth]{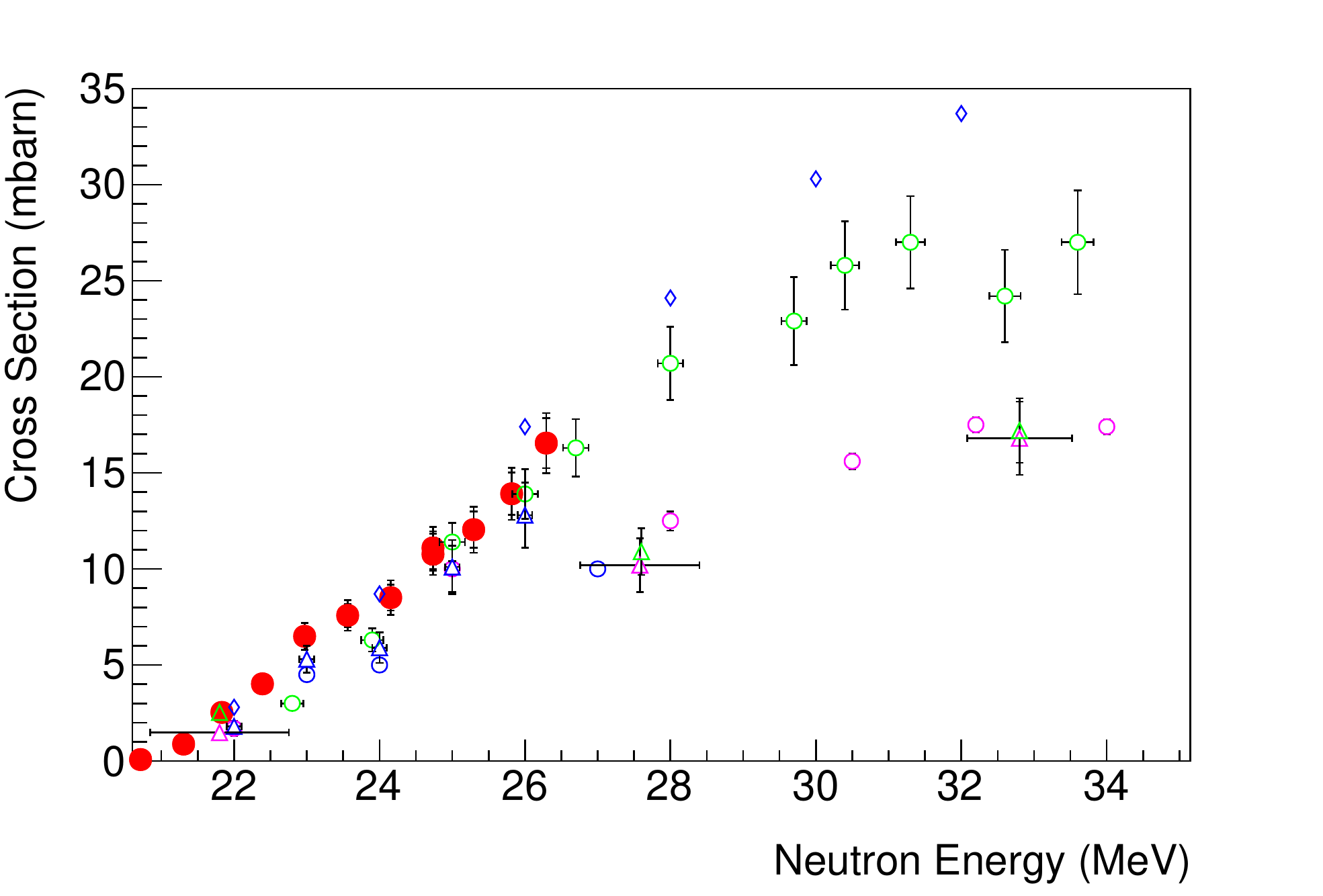}% Here is how to import
		%EPS art
		\caption{\label{fig:cross_sections} Cross sections for the $^{12}$C(n,
			2n)$^{11}$C reaction near threshold. The empty symbols are previously published
			data: Brolley et al. (blue circles) \cite{brolley1952},  Brill et al. (pink
			circles) \cite{brill1961}, Anders et al. (green circles) \cite{anders1981},
			Welch et al. (blue triangles) \cite{welch1981},  Soewarsono et al. (pink
			triangles) \cite{soewarsono1992},  Uno et al. (green triangles) \cite{uno1996},
			and optical model calcualtions of Dimbylow (blue diamonds) \cite{dimbylow1980}. The solid red symbols are from
			this experiment using the polyethylene target and both NaI detectors in
			coincidence; the larger of the associated error bars indicates the increase in overall uncertainty when the uncertainty in incident neutron energy is included.  }
	\end{figure}
	
	It is difficult to see what might be causing the results for these experiments
	to fall into two distinct bands.   Table~\ref{tab:prev_measure} summarizes the
	published  essential features of each previous experiment.  There does not seem
	to be any obvious division between the bands on the basis of technique, type of
	neutron source, type of target, method of neutron flux determination or type of
	activation measurement.  
	
	\begin{table*}[]
		\caption{\label{tab:prev_measure} Previous measurements of the $^{12}$C(n,2n)
			cross section in the energy range 20-30 MeV.  }
		%	\resizebox{\textwidth}{!}{%
		\begin{ruledtabular}
			%	\begin{tabular}{lllllll}
			\linespread{0.8}                    %   this decrease the vertical spacing
			%between lines
			\small
			\rmfamily  
			\setlength\extrarowheight{12pt}
			\begin{tabular}{ >{\raggedright}p{0.7cm}  >{\raggedright}p{3cm}
					>{\raggedright}p{2.5cm}  >{\raggedright}p{2.5cm}  >{\raggedright}p{1.5cm}
					>{\raggedright}p{2cm}  p{2.5cm} }
				
				Year & Experiment & Accelerator & Neutron Source & Target & Neutron Flux &
				Activation Measurement \\
				\hline
				
				1952 & Brolley et al. \cite{brolley1952} & 10.5 MeV deuterons from cyclotron &
				$^3$H(d,n)  gas cell, neutron energy selected by angle & Poly- ethylene foils &
				Calculated from $^3$H(d,n) cross section & Geiger counter \newline calibrated
				with \newline RaD+E \\
				1961 & Brill et al. \cite{brill1961} & 20 MeV deuterons from cyclotron, Pt
				foil degrader & $^3$H(d,n) (Zr foil) and $^2$H(d,n) (gas) & Carbon & TOF
				energy/angle distribution & Geiger counter \newline calibrated with \newline
				$^{197}$Au \\
				1981 & Anders et al. \cite{anders1981} & 7-16 MeV deuterons from cyclotron, Be
				foil degrader  & $^3$H(d,n)  (Ti foil) & Reactor graphite & Stilbene crystal
				recoil proton spect. & Annihilation $\gamma$-$\gamma$ coincidence using NaI
				detectors \\
				1981 & Welch et al. \cite{welch1981} & Tandem Van de Graaff & $^3$H(d,n) &
				Natural carbon & No information available. & Ge(Li) Detector calibrated with
				\newline $^{22}$Na \\
				1992 \newline 1996 & Soewarsono et al. \cite{soewarsono1992} \newline Uno et
				al. \cite{uno1996} & 20-40 MeV protons from cyclotron & $^7$Li(p,n) 
				quasi-monoenergetic & $^7$Li on graphite  & From activation of Li target &  HPGe
				detector, Li in/out subtraction \\
				
			\end{tabular}
		\end{ruledtabular}
	\end{table*}

	Since it is not clear why the previous measurements disagree, the present
	experiment was designed to reduce or eliminate possible sources of systematic
	uncertainty that may have affected previous results.  In the present experiment,
	monoenergetic neutrons produced using the $^3$H(d,n)$^4$He reaction were allowed
	to strike carbon-containing targets of polyethylene and graphite.  When these
	neutrons induced the $^{12}$C(n, 2n)$^{11}$C reaction, $^{11}$C nuclei were
	produced, which later decayed via $\beta^+$ emission with a half-life of
	$20.364\pm0.014$  minutes \cite{kelley2012}.  After an activation period, the
	targets were removed to counting stations, where both the singles and
	coincidence rates of 511 keV gamma rays resulting from positron annihilation
	were used to determine the number of $^{11}$C nuclei present.  In order to
	determine the neutron flux, protons from neutron-proton elastic scattering were
	simultaneously counted in a $\Delta$E-E detector telescope.
	
	The present experiment has a number of advantages:
	\begin{enumerate}
		\item  The electrostatic accelerator provides an extremely stable and nearly
		mono energetic deuteron beam, which, when used with a very thin titanium tritide
		target gives intrinsic neutron energy spread of less than about 0.2 MeV in the
		20 to 30 MeV neutron range.
		\item  By  using a recoil proton telescope with $\Delta$E-E silicon detectors,
		the present experiment has  ability to identify and select only $^{1}$H(n,p) elastic
		recoil protons of the correct energy.  Since the solid angle is well defined and
		the intrinsic efficiency is nearly 100\% for the silicon detectors, an absolute
		determination of the neutron flux is possible using well measured np elastic
		scattering cross sections.
		\item Two targets, graphite and the polyethylene proton production target,
		irradiated and counted simultaneously, allow a consistency check. Moreover,
		since the polyethylene activation target is also the proton production target
		for the telescope, problems that would result from the neutron flux being
		measured in a different place than the activation target are eliminated.
		\item Counting the activated targets by  placing them between the circular faces
		of two cylindrical
		matched  NaI crystals gives maximum solid angle and therefore maximum absolute
		efficiency for counting the relatively small number of $^{11}$C decays, thereby
		reducing counting uncertainty and the effect of unwanted background.  Requiring
		a coincidence rather than using a single detector  eliminates  most background
		511 keV gamma ray events that do not come from $^{11}$C decay.
		\item  A careful study was made of the absolute full-peak efficiency of the
		counting system.   A Monte Carlo code was developed to calculate this efficiency
		for both singles and coincidence mode geometries, the results of which were
		validated by comparisons to a number of ancillary experiments.  This allowed a
		consistency check to be made by simultaneously measuring the cross sections
		using both the coincidence and singles count rates.
	\end{enumerate}
	
	\section{Description of the Experiment} 
	
\begin{figure}
		\includegraphics[width=\linewidth]{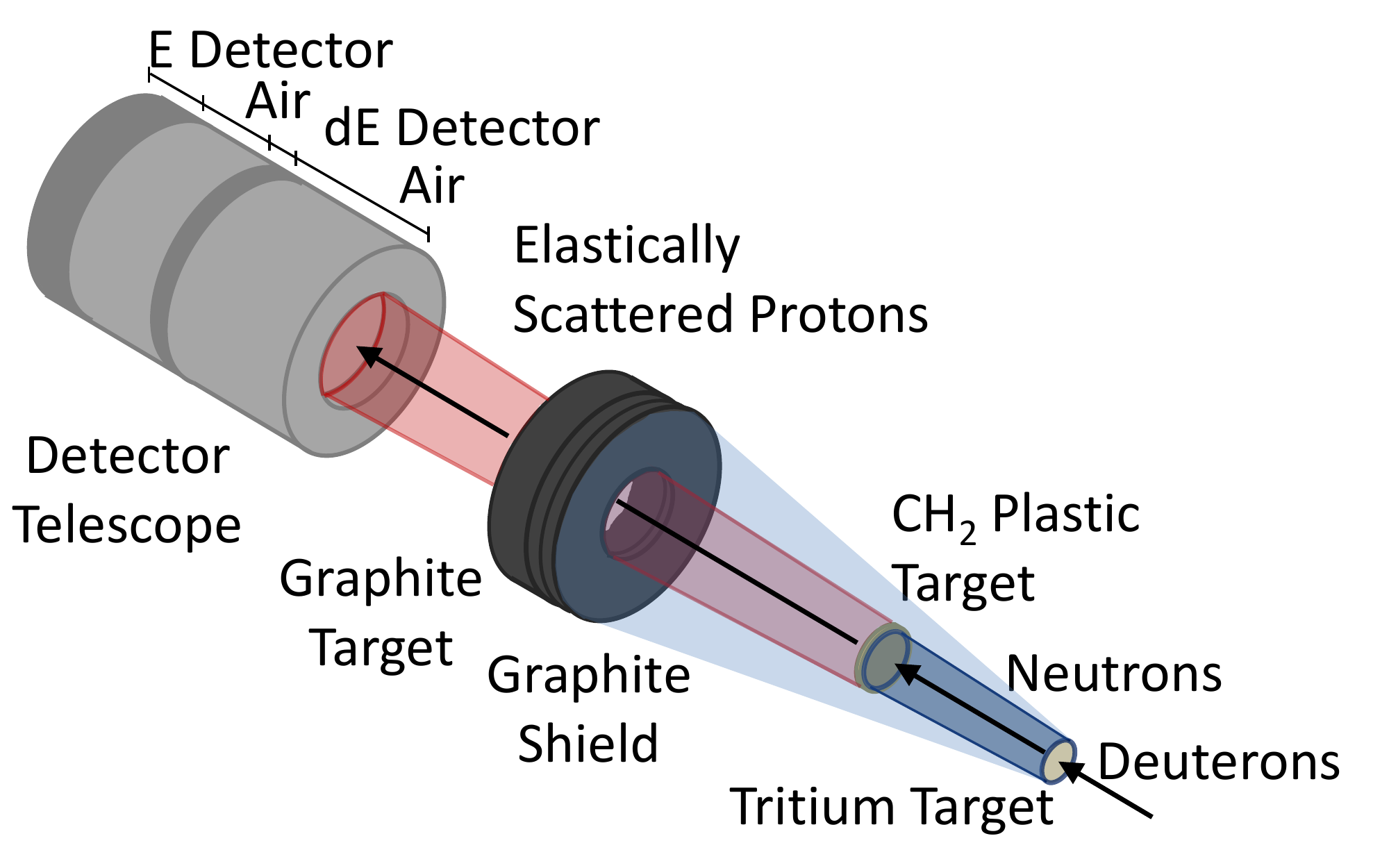}% Here is how to import EPS art
		\caption{\label{fig:setup} The experimental setup for activating targets. 
			Deuterons traveling down the beam pipe were collimated, then struck the tritium
			target to produce neutrons.  These neutrons activated the graphite and
			polyethylene targets via the $^{12}$C(n,2n) reaction.  The polyethylene target
			also acted as the converter for the recoil proton telescope.  The recoiling
			protons were identified and counted using a dE-E telescope. }
	\end{figure}

	The cross sections were measured for energies between about 19.7 and 26.4 MeV
	using the 4.5 MV Tandem Pelletron 
	electrostatic accelerator at Ohio University.
	As shown in Fig.~\ref{fig:setup}, deuterons were accelerated to energies between
	3.1 and 9.1 MeV and allowed to strike a 472.86 GBq tritium target that was perpendicular to the beam and located at the end of the beam pipe just upstream of an aluminum end window.  Analyzing magnet
	image slits restricted the spread in deuteron beam energy to about 5 keV. 
	Deuteron beam currents were typically between 0.8 and 1.0 $\mu$A.  The tritium
	was deposited as titanium tritide on a 49 mm diameter 1 mm thick OFHC copper
	substrate at a density of about 2000 $\mu$g/cm$^2$ over a circular active area
	of 30 mm diameter.   The end window where the target was mounted was cooled by a stream of air. The target assembly was attached to the beamline with a bellows so that the target could be rotated in a circular path in a plane perpendicular the beam direction. The radius of rotation was approximately 1 cm and the period of rotation was about 0.3 s which spread the beam heating and target sputtering over a larger area on the target without compromising the geometry of the experiment. Before striking the target, the deuteron beam was defocused by a
	pair of quadrupole magnets located 275 and 315 cm upstream, and allowed to pass
	through a 1.27 cm diameter collimator 45 cm upstream of the target.  This was to ensure that the beam spot on the target was relatively large, uniform, and in a known and reproducible location.
	These characteristics minimized local heating of the target and were also needed for simulations described below.

	A 50 cm long steel optical bench with modified positioners  was used to  hold
	the detectors and targets in fixed positions at an angle of 0$^\circ$ with
	respect to the beamline. The targets were aligned to the optical center of the beam line using the procedure discussed below.
	
	Neutrons leaving the tritium could strike a 1.64 mm
	thick, 2.54 cm diameter high-density polyethylene target with its upstream face
	located 7.0 cm from the tritium target and a 7.62 cm diameter 0.89 cm thick disk
	of high purity graphite with a 17.46 mm hole drilled in through its center with
	its upstream face 14.4 cm from the tritium target.  In contact with the upstream
	face of the graphite target, a pair of graphite disks stopped any protons
	scattered from the polyethylene from reaching the graphite target.  The total
	thickness of this graphite shield was 3.85 mm, with the other dimensions being
	the same as the graphite target disk.  The mass densities of the polyethylene and graphite targets were measured to be  $0.957 \pm 0.008$ and $1.842 \pm 0.012$ g/cm$^2$ respectively. 
	
	A proton telescope consisting of a 300 $\mu$m thick, 150 mm$^2$ ion implanted
	silicon dE detector and a 5000 $\mu$m thick, 200 mm$^2$ drifted-lithium silicon
	E detector was placed behind the hole in the graphite target, so that protons
	coming from the polyethylene could be viewed.  The entrance of the dE detector
	was covered by a 0.005 mm thick aluminum foil to keep out ambient light.  The
	entire detector assembly was housed in an aluminum tube with wall thickness of
	about 2.9 mm and diameter of 3.47 cm.  Preamplifiers and spectroscopy amplifiers
	located near the detectors in the experimental hall sent pulses to a FastComTech
	MPA-3 multiparameter system which digitized and recorded the pulse heights and
	timing.  The system also recorded the deuteron beam current.
	
\begin{figure}
		\includegraphics[width=\linewidth]{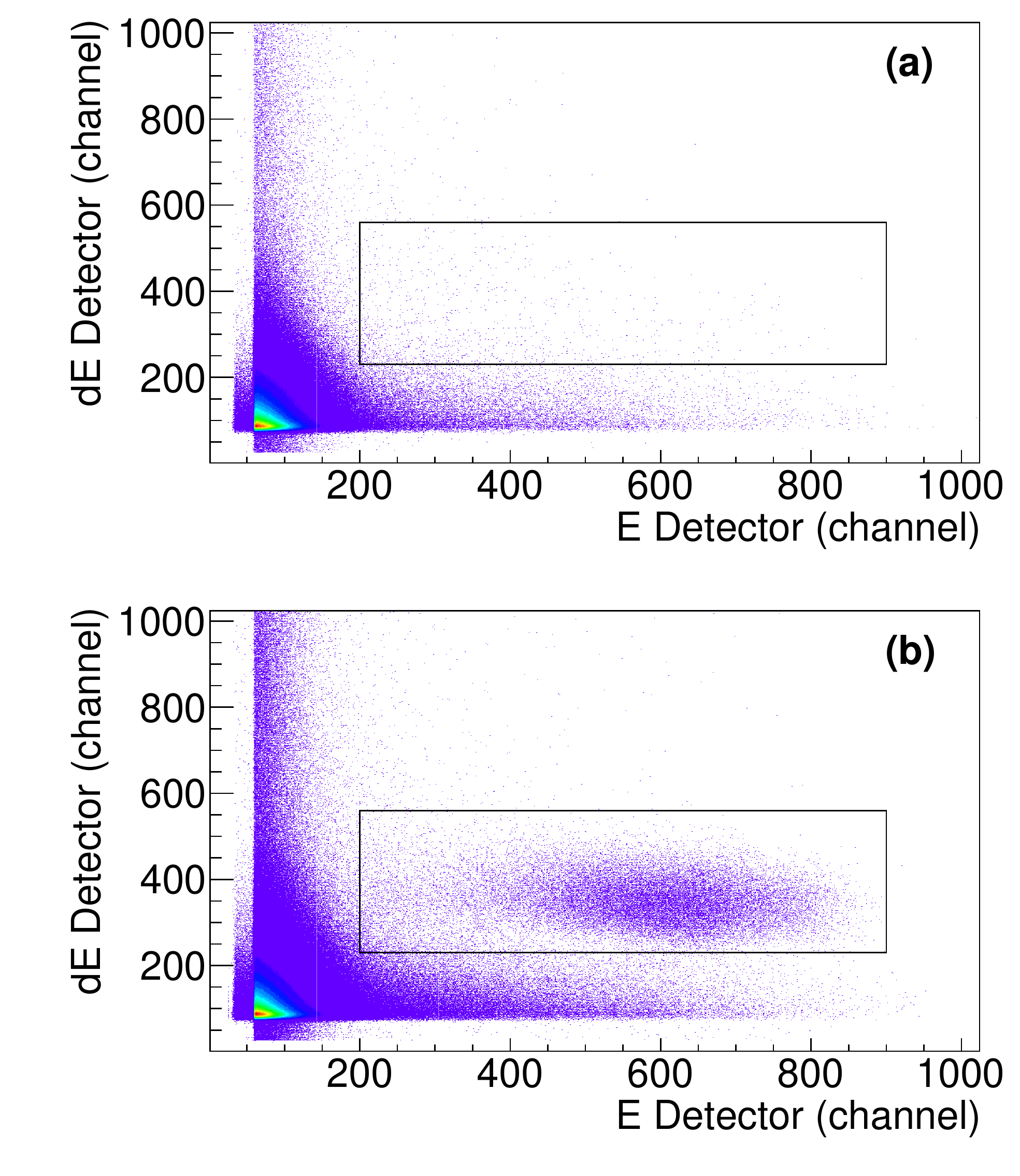}% Here is how to
		%import EPS art
		\caption{\label{fig:telescope_histogram} (b) Histogram of dE versus E for
			23.7 MeV neutrons striking the polyethylene target.  The proton island is
			clearly visible.  The marked region-of-interest (black) indicates the
			elastically scattered protons. (a) Same, but with the polyethylene target
			removed. }
	\end{figure}

	Fig.~\ref{fig:telescope_histogram} shows 2D histograms of the pulse height in
	the dE detector versus the E detector. The $^{1}$H(n,p) protons elastically scattered from the polyethylene
	can be easily identified by their energy loss in the two detectors.  At 26.4 MeV
	the background count rate was about 3\% of the rate with the polyethylene target
	in place, which was approximately 3.5 protons/sec with a beam current of about 1
	$\mu$A.  Over the course of the experiment, radiation damage caused the width of
	the proton peak in the E detector to gradually increase along with the leakage
	current.   
	
	A special circuit provided a separate  count of the number of coincidence
	events, and gated the individual detector pulses that were input to
	analog-to-digital converters (ADCs) that were part of the  FAST ComTec MPA-3
	multiparameter system.   The live time, which was typically about 98\%, was confirmed by comparing the number of coincidence events recorded
	by the computer with the number counted by a separate hardware circuit. 
	
	A 12.7 cm diameter, 5.08 cm thick NE-213 liquid scintillator neutron monitor was located 300 cm from
	the tritium target at an angle of a 71.4$^\circ$ to beam left.  In order to
	identify neutron pulses from gamma ray signals in the monitor, a pulse shape
	analysis was made using a Mesytec MPD4 pulse shape discriminator module.  These
	signals were also recorded by the MPA-3 multiparameter system.
	
	% \begin{figure}
	% 	\includegraphics[width=\linewidth]{fig_circuit}% Here is how to import EPS
	%art
	% 	\caption{\label{fig:circuit} Circuit diagram for external coincide gating
	%circuit.  Pulses from the dE and E silicon detectors were  gated and required
	%to be in coincidence before they were recorderd by the ADCs.  This circuit
	%allowed a hardware count of the number of coincidence events permitting an
	%independent determination of the computer dead time. }
	% \end{figure}
	
\begin{figure}
		\includegraphics[width=\linewidth]{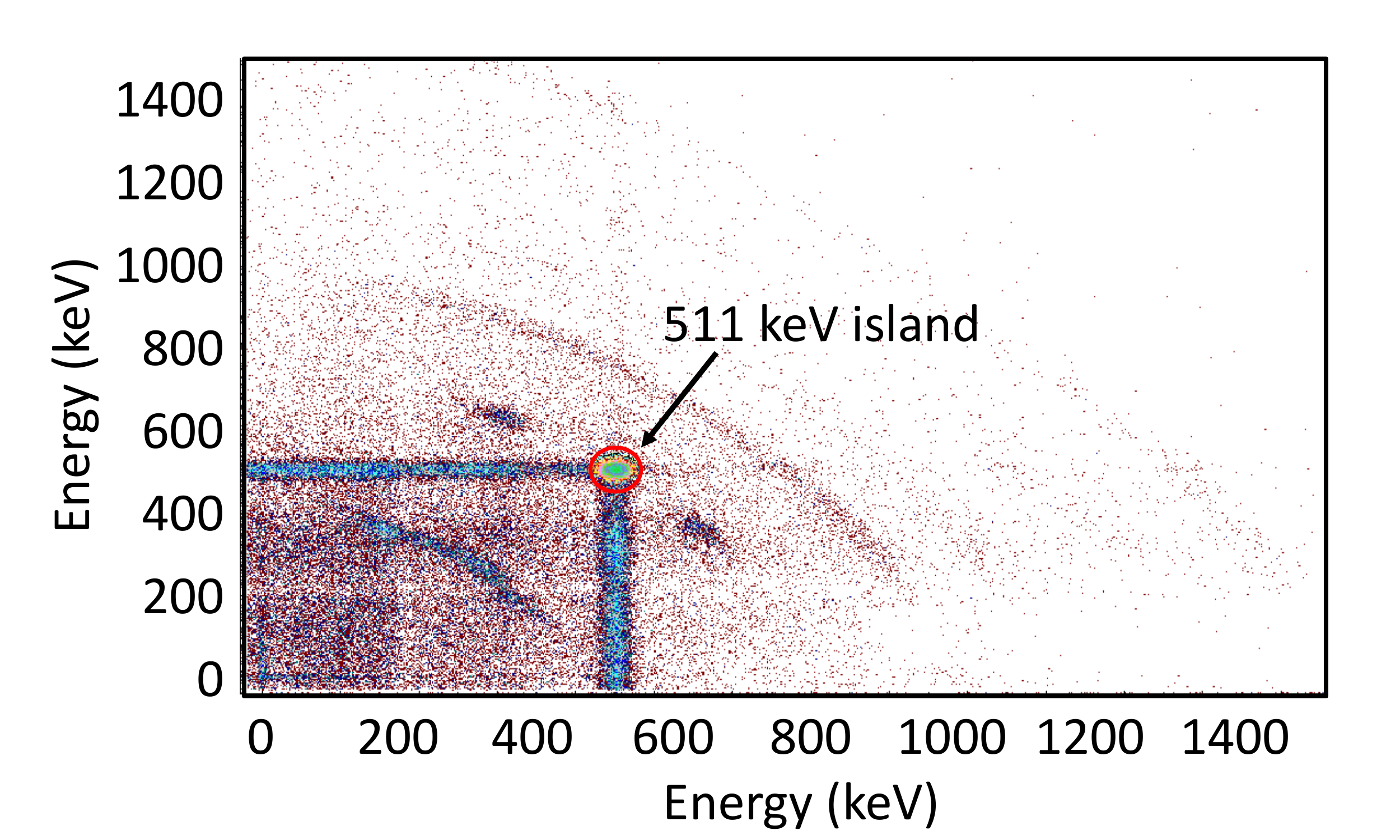}% Here is how to import EPS
		%art
		\caption{\label{fig:counting} A 2D histogram showing the pulse height in each
			NaI detector for coincidence events.  The events in the large coincidence peak were counted for 50
			second intervals to produce the $^{11}$B growth curve. }
	\end{figure}

	Activated targets were counted at three counting stations located in a room far
	from the accelerator target area to reduce background counts.  After each was
	activated simultaneously, the graphite target disks, shields and polyethylene
	targets were placed between pairs of 7.62 cm diameter by 7.62 cm thick NaI
	detectors. Pairs of detectors and the graphite targets were held with their axes aligned inside almost equal diameter acrylic tubes.  The polyethylene targets, which were much smaller in diameter, were affixed with adhesive tape to the center of one of the detectors, which was marked. 
	Pulses from all of these detectors were digitized by a FAST ComTec
	MPA-4 system, which recorded the pulse heights and timing information. 
	Coincidence events consisting of two back-to-back 0.511 keV gamma rays from
	positron annihilation were selected (as shown in Fig.~\ref{fig:counting}) and
	counted as a function of time.   This allowed the growth curve of $^{11}$B to be
	measured and fit in order to determine the number of $^{11}$C nuclei present.
	The gamma rays from the graphite shields were also counted in a separate station
	consisting of two high-purity germanium detectors to look for activation due to
	contaminants in the graphite.  
	
	\section{Procedure}
	
	The silicon detector telescope and targets were positioned at 0$^\circ$ to the
	beam line using a theodolite that had been previously aligned with the
	collimator in the beam pipe and monuments in the target room.  
	
	To ensure the polyethylene target holder was centered and the neutron
	distribution was axially symmetric, a duplicate target was cut in half and the
	proton count rate was measured to approximately 2\% statistical uncertainty with
	the half target in each of four positions, top, bottom, left and right.  The
	target position was changed by simply rotating it in the holder, and the ratio
	of protons to integrated charge on the tritium target, which should be constant
	for perfect axial symmetry, was found to vary by less than 10\%.  This small deviation from a uniform neutron distribution on the polythene target would only result in a small correction to the already small extended target correction described below.      
	
	For each energy setting, prior to activating the targets, a study was made of
	the beam defocussing and positioning of the beam along the axis of the beam
	pipe.   This step is important because the distribution of the beam on the target must be input to the simulation of the experiment described below. 
	Experimentally, it was observed that steering the beam off of the central axis of the collimator would change the number of protons per integrated charge on the tritium target and the ratio of protons to neutrons detected in the neutron monitor. In order to place the beam on the central axis of the quadrupoles, the steering of the beam was adjusted such that the ratio of detected protons to neutrons was minimally sensitive to the the quadupole current (less than a 5\% change for a quadrupole current change of 25\%). The deuteron beam was also monitored with a beam profiler upstream from the target. These procedures ensured that the beam was very closely aligned with the optical beam axis and centered on the target.

	Following these quadrupole tests, a shield, graphite target and a polyethylene
	target were placed in the target holders to be activated for about 1.5 hours,
	during which time the proton telescope pulses were recorded.  Three identical
	sets of targets were available to be used consecutively in order to allow enough
	time for any longer lived activated contaminants to decay between uses, although
	no contaminants were detected.  When the deuteron beam was stopped, after
	sufficient time to allow the room radiation dose rate to fall to an acceptable
	level, the targets were hand-carried to the counting room and placed in the
	counting stations.  The time between when irradiation stopped and when target
	counting commenced was  typically 4-5 minutes.  Each sample was then counted for
	approximately 2 hours, binned into 50 second intervals.  The dead time for each
	time bin was recorded and used to correct the growth curve.   
	
	The background proton count rate was measured at each energy setting by removing
	the polyethylene target and counting for approximately 30 minutes.  A separate
	graphite disk target and shield was used exclusively for this purpose.
	
	\section{Analysis}
	\subsection{Overview}
	The $^{11}$C decays in the activated polyethylene and graphite targets were
	counted using pairs of NaI detectors ``sandwiching" each target, capable of
	counting the 511 keV gamma rays from positron annihilation in both singles and
	coincidence modes simultaneously.   This allowed the $^{11}$B growth curves from
	$^{11}$C\ $\rightarrow\ ^{11}$B$\ +\ $e$^+ +\nu$  to be measured for both
	singles and coincidence events, and fit with the exponential growth function
	\begin{equation}
	R(t_c)=R_0 (1-e^{-\lambda t_c})+At_c+B
	\label{eq:fit}
	\end{equation}
	where $R(t_c)$ is the sum of all the positron annihilation events counted up to
	time $t_c$, $R_0$ is the total number of detectable $^{11}$C decays,
	$\lambda=  20.364 \pm 0.014$ min  \cite{kelley2012} is the decay constant for $^{11}$C
	and $At_c+B$ is the integral of the constant rate of background events.  The number of counts $N_0$ that would be obtained if counting began immediately at the end of activation is
	\begin{equation}
	N_0 =  { R_0 e^{\lambda t_{\textbf{trans}}}}
	\label{eq:number_det_tot}
	\end{equation}
	where $t_{\text{trans}}$ is the time between the end of activation and the start
	of counting, in other words, the time required to transfer the samples to the
	counting station.  The total number of $^{11}$C nuclei formed in the target is
	\begin{equation}
	N_{^{11}\text{C}} = 	\frac { N_0} {\epsilon}
	\label{eq:number_of_11C}
	\end{equation}
	where  $\epsilon$ is either the absolute full-peak coincidence or singles
	efficiency, depending on how the growth curve was generated.
	
	Cross sections $\sigma$ for $^{12}$C(n,2n) were extracted using
the above quantities as well as the background-subtracted rate of elastically scattered
	protons detected, $N_p$, and the activation time, $t$, since
	\begin{equation}
	\sigma =  
	\frac{ N_0 } {\epsilon}
	\frac{1}{T_C}
	\frac {\lambda} {1-e^{-\lambda t}} 
	\left( \frac{N_p} {N_n} \right) 
	\frac {1} {N_p}
	\label{eq:cross_section}
	\end{equation}
	where $T_C$ is the
	target thickness in terms of carbon nuclei (carbon nuclei/cm$^2$).  The quantity
	$\left(  {N_p}/{N_n} \right)$ is the ratio of elastically scattered protons
	detected, $N_p$, to the number of neutrons  striking the polyethylene or graphite target, $N_n$. 
	This ratio was calculated numerically for the experiment geometry for each target using the known
	$^3$H(d,n)$^4$He \cite{drosg2000}   and $^1$H(n,p)n elastic scattering
	\cite{Stoks1993,Stoks1994}  cross sections.  
	
	\subsection{Growth Curves}
	To determine $R_0$, Eq.~(\ref{eq:fit}) was fit to the growth curves for the polyethylene and graphite
	targets using the c++ ROOT \cite{Brun1997} TMinuit class implementation of the
	Minuit package \cite{james1998}.   Fits were made using singles events from each
	NaI detector individually and also for coincidence events, resulting in six
	semi-independent measurements of the cross section. For the polyethylene target,
	which is much thinner than the graphite,  at the highest incident neutron energy
	setting about 7000 $^{11}$C decays were counted over the 2 hour period.  
	
\begin{figure}
		\includegraphics[width=\linewidth]{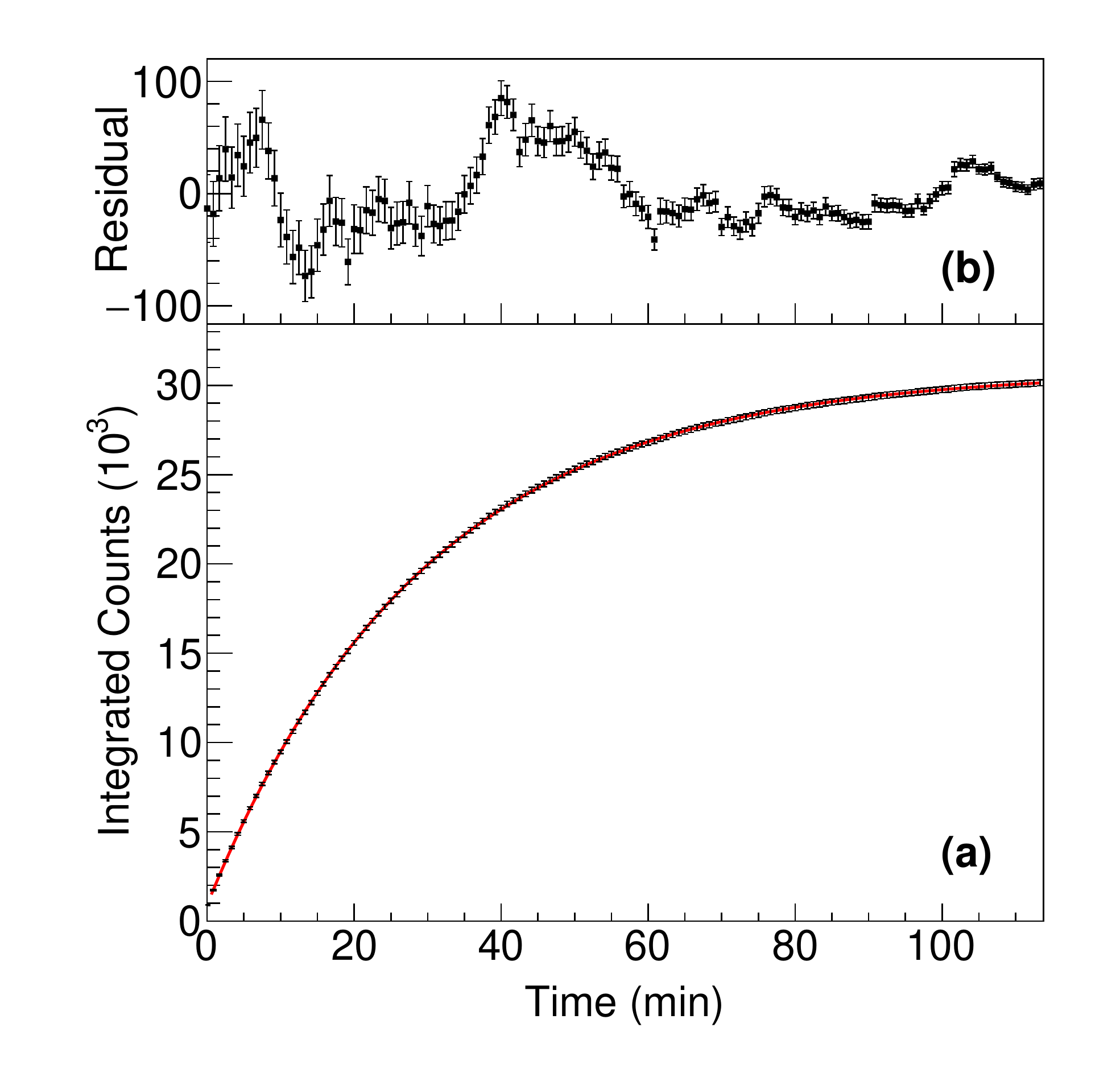}% Here is how to import EPS art
		\caption{\label{fig:growth}  (a) Fit of Eq.~(\ref{eq:fit}) to the coincidence growth curve and (b) residuals   for
			the graphite target activated by 26.3 MeV neutrons. The uncertainties shown were calculated using the integrated number of counts (a) and the incremental number for each time bin $\sqrt{N_i-N_{i-1}}$ (b).  }
	\end{figure}

	Fig.~\ref{fig:growth} shows a typical fit of  Eq.~(\ref{eq:fit}) to a
	coincidence growth curve.  These growth curves were created by
	integrating the number of events in the 511 keV peak up to time $t_c$, and
	plotting the integral as a function of $t_c$.  The exponential nature of the
	growth curve was clear for coincidence events, but because of the large number of
	background events relative to the number of $^{11}$C decays, the singles growth
	curves for the thin polyethylene targets were nearly a straight lines. 
	Nevertheless, the value of $R_0$  could still be extracted from these fits and
	used to determine the cross section using Eq.~(\ref{eq:cross_section}) , albeit with a larger
	uncertainty.

	\subsection{Determination of $\left( {N_p} / {N_n} \right) $}
	
	In order to figure out the cross section for the  $^{12}$C(n, 2n)$^{11}$C 
	reaction using  Eq.~(\ref{eq:cross_section}), the incident neutron flux $N_n$ striking each target was determined from the number of protons $N_p$ using 
	\begin{equation}
	N_n =  \frac {1} {\left( \frac{N_p} {N_n} \right) }
	 {N_p}
	\label{eq:n_flux}
	\end{equation}
	where the quantity in parenthesis, $\left( {N_p} / {N_n} \right) $, the ratio of  the number of protons detected to neutrons striking the target was calculated purely from the experiment geometry.
	
	To calculate the ratio $\left( {N_p} / {N_n} \right) $ for each target,  several simplifying assumptions were
	made to which corrections were later applied.   Assume that the tritium target
	is a point source of monoenergetic neutrons, isotropic in the lab frame, with
	flux $N$ (neutrons/sec/sr).  In that case, the number of neutrons/sec ($N_n$) striking
	the polyethylene ($N_n=N_\text{CH2}$) and graphite ($N_n=N_G$) targets would be given by $N_\text{CH2}
	= N \Omega_\text{CH2}$ and $N_G = N \Omega_G$ where $\Omega_\text{CH2}$ and $\Omega_G$ are
	the solid angles of the plastic and graphite targets, respectively, if the
	effect of the finite thickness of the targets on solid angle is neglected.  The
	number of protons/sec ($N_p$) detected by the proton telescope can be  obtained
	if the polyethylene target is then treated as a point source for the purpose of
	scattering protons into the proton telescope.  Assuming that the cross section
	for $^{1}$H(n,p) elastic scattering, $\sigma_{np}\left( \psi_{np}, E_{n} \right)$, which
	depends on the scattering angle, $\psi_{np}$, and the incident neutron energy,
	$E_n$, is roughly constant over the angles subtended by the targets, and is
	equal to the cross section at $0^\circ$ at the nominal neutron energy, yields
	\begin{equation}
	N_p = \sigma_{np}(0^\circ) T_H N_n \Omega_p
	\label{eq:protons}
	\end{equation}
	where $T_H$ is the thickness (hydrogen nuclei/area) of the polyethylene target,
	and $\Omega_p$ is the solid angle of the proton telescope.  In this simple
	approximation, therefore, the calculated ratio of the rates for protons detected
	by the proton telescope to neutrons hitting the polyethylene target is
	\begin{equation}
	\left( \frac{N_p} {N_n} \right)_\text{CH2} = \frac{N_p} {N_\text{CH2}}= \sigma_{np}(0^\circ) T_H \Omega_p
	\label{eq:simple_ratio_CH2}
	\end{equation}
	and for the graphite target
	\begin{equation}
	\left( \frac{N_p} {N_n} \right)_G = \frac{N_p} {N_G}= \sigma_{np}(0^\circ) T_H \frac {\Omega_\text{CH2} \Omega_p}{\Omega_G}
	\label{eq:simple_ratio_G}.
	\end{equation}
	Clearly, the approximation that the tritium and polyethylene targets can be
	treated as point sources is incorrect, and a more correct solution must include
	the fact that the neutrons leaving the tritium target can have a range of angles
	and still reach the polyethylene and graphite targets, and that the energies and
	cross sections for these neutrons depends on the neutron angle.  Moreover, the
	protons coming from the polyethylene, which is actually an extended source, can
	also have a range of angles and still strike the proton telescope, and the cross
	sections and energies of the protons reaching the telescope depends on the
	proton angle.  
	
\begin{figure*}
		\includegraphics[width=\linewidth]{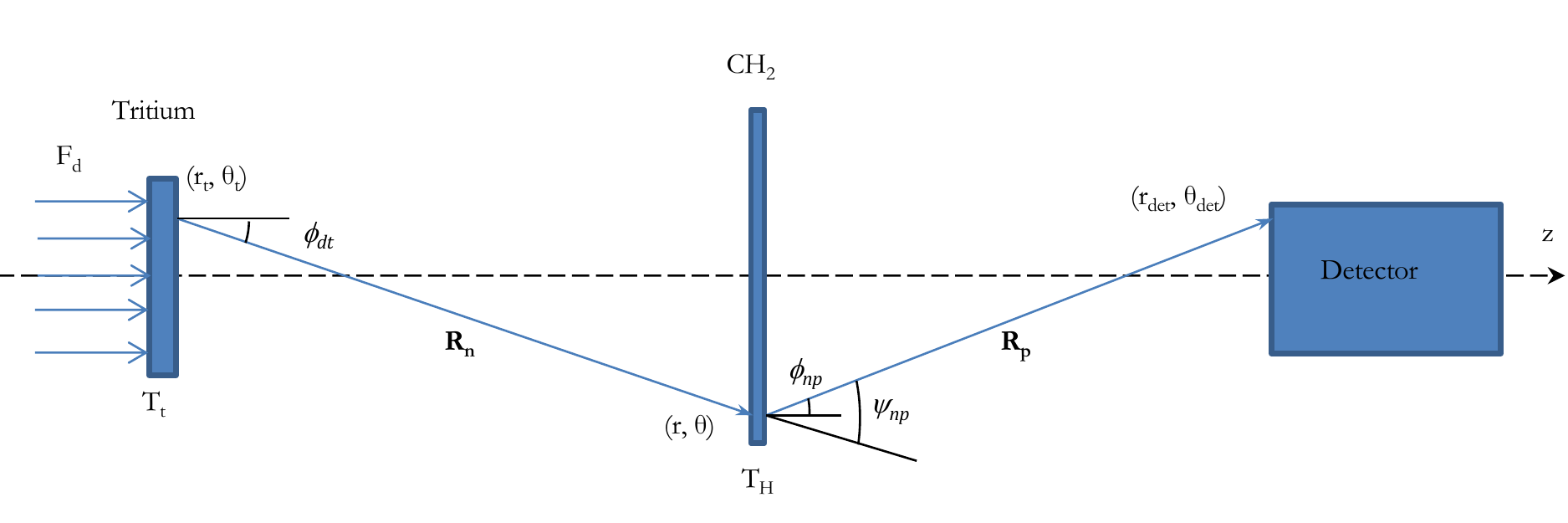}% Here is how to import EPS
		%art
		\caption{\label{fig:integrate} Schematic diagram showing quantities used to
			calculate the  $	\left( {N_p} / {N_n} \right) $ proton to neutron ratio for
			extended targets. A neutron can travel from an infinitesimal area element at
			polar coordinates $( r_t, \theta_t )$ on the tritium target to an area element
			at polar coordinates $(r,\theta)$ on the polyethylene target along vector
			$\mathbf{R_n}$. A proton can then travel from there to the area element at polar
			coordinates $(r_{det},\theta_{det} )$ on the detector along vector
			$\mathbf{R_p}$. The total number of neutrons or protons were determined by
			integrating over the surfaces of the involved targets or detector. The $z$
			coordinate axis is along the center of the beam line. }
	\end{figure*}

	A number of corrections were applied to this simple calculation of the  $	\left(
	{N_p} / {N_n} \right) $ proton to neutron ratio. The first correction accounts
	for the fact that the targets were not point sources but actually extended
	targets. The $^{1}$H(n,p) elastic scattering and DT fusion cross sections both depend on
	the scattering angle, as do the energies of the scattered neutrons and protons.
	To account for these effects, the surface of each target and the silicon
	detector were divided into infinitesimally small area elements, each of which
	was then treated as a point source. The total number of protons or neutrons
	hitting a target or detector was determined by integrating over the surface area
	of each target, as depicted in Fig.~\ref{fig:integrate}. 
	
	The rate of neutrons hitting the polyethylene target is
	\begin{eqnarray}
	N_\text{CH2} &=& \int_{0}^{2\pi } \int_{0}^{R} \int_{0}^{2\pi} \int_{0}^{R_{t}}
	\sigma _{dt} ( \phi_{dt},E_{d} ) \nonumber \\
	& & \times F_{d} T_{t}  r_{t} dr_{t} d\theta_{t} 
	\frac {\cos \phi _{dt}} {R_{n}^{2}} r dr d\theta 
	\label{eq:n_polytarget}.
	\end{eqnarray}
	where  $R_t$ is the radius of the beam spot on the tritium target (0.635 cm),
 $R$ is the radius of the polyethylene target, the incident deuteron energy is $E_d$, the deuteron flux
	(deuterons/area/time) is $F_d$, the thickness of the tritium target ($^3$H nuclei
	per unit area) is $T_t$, and the surface area of the tritium target is $A_t$. 
	The cross section for DT fusion at the neutron angle $\phi_{dt}$ for incident
	deuteron energy $E_d$ is $\sigma_{dt}(\phi_{dt},E_d )$ where $\phi _{dt}=\cos
	^{-1} (  {\mathbf{R_{n}} \cdot \mathbf{\hat {z}}} / {R_{n}} )$.  
	
	The number of neutrons impacting the graphite target ($N_G$) was calculated using the same formula by
	integrating over the surface area of the graphite target rather than the
	polyethylene target, remembering to include the central hole.
	
	The rate of protons striking  the proton telescope due to $^{1}$H(n,p) elastic scattering of
	DT neutrons from protons in the plastic target is given by
	\begin{eqnarray}
	N_{p} &=& \int_{0}^{2\pi } \int_{0}^{R_{det}} \int_{0}^{2\pi} \int_{0}^{R}
	\int_{0}^{2\pi } \int_{0}^{R_{t}}
	\sigma _{np} \left( \psi_{np}, E_{n}(\phi _{dt},E_{d}) \right) \nonumber \\
	& & \times \sigma_{dt}(\phi_{dt},E_{d} ) 
	F_{d} T_{t} \frac {T_{H}} {\cos \phi_{dt}}
	r_{t} dr_{t} d\theta _{t} 
	\frac{\cos \phi_{dt}} {R_{n}^{2}} r dr d\theta \nonumber \\
	& & \times  \frac{\cos \phi _{np}} {R_{p}^{2}} r_{det} dr_{det} d\theta_{det}
	\label{eq:p_telescope}.
	\end{eqnarray}
	where $R_{det}$ is the radius of the proton detector, $E_{n}(\phi _{dt},E_{d})$
	is the DT neutron energy, which depends on the deuteron energy and DT neutron
	angle  $\phi_{dt}$, and $\psi_{np}$ is the np scattering angle, given by   $\psi
	_{np}=\cos ^{-1} ( \mathbf {R_{n}} \cdot \mathbf{R_{p}} / {R_{n}R_{p}} )$.  
	
	The hole through the center of the graphite target was slightly too small for
	all of the protons to pass through unobstructed, thereby causing a collimating
	effect. To correct this, if in the integral any proton path intersected the
	graphite disk it was excluded from the integral. This changed $N_p$ by less than
	0.5\%.
	
	A c++ program was written to evaluate these integrals numerically using the
	rectangle method. Each integral was divided into 150 steps, the number of steps
	chosen based on the rate of convergence to give around 1\% uncertainty in the
	proton-to-neutron ratio. The scattering angles for the DT fusion and $^{1}$H(n,p) elastic
	scattering reactions for each neutron and proton path were determined for each
	step. No neutron scattered at an angle of greater than around $17^\circ$ was
	able to hit the graphite target. No proton scattered at an angle greater than
	around $25^\circ$ could hit the detector.  Since these extreme angles were very
	unlikely, most paths had cross sections fairly close to the  $0^\circ$
	approximation. 
	
	The cross section measurement was insensitive to the Ti:T ratio and the overall tritium activity of the tritium target, since  the neutron flux was measured directly using the proton telescope.  However, the neutrons striking the graphite and polyethylene targets are
	slightly reduced in energy and have a broadened energy distribution as a result
	of the deuterons losing energy prior to interacting with the tritium, and the
	resulting neutrons leaving at angles greater than $0^\circ$. The thickness of the
	tritium target was also divided into 150 steps, and for each step in thickness
	and angle the energy of the neutrons striking the targets was calculated.  The
	calculated neutron spectra were used to correct the nominal neutron energies. 
	To do this, for neutron energies above about 21.5 MeV, where the  $^{12}$C(n,
	2n)$^{11}$C cross sections are large enough to allow, a quadratic polynomial was
	fit to the  measured $^{12}$C(n, 2n)$^{11}$C cross sections as a function of 
	uncorrected nominal  neutron energy. This was used with the calculated neutron
	energy spectrum to predict the expected $^{11}$C distribution in each target. 
	Then the total $^{11}$C in the target  and the preliminary cross section fit
	were used  to determine the corrected neutron energy -- that is, if all the
	neutrons had this energy, they would give the same number of $^{11}$C nuclei  as
	the actual neutron distribution. This process resulted in a maximum downward
	shift in neutron energy of less than 1.3\% (0.28 MeV) which occurred for 5.57
	MeV deuterons.  The FWHM of the energy neutron energy distribution striking the
	targets was typically about 0.3 MeV.  For deuteron energies below 21.5 MeV,
	where the  $^{12}$C(n, 2n)$^{11}$C cross section is essentially zero, the average was used to
	estimate the corrected neutron energy.
	
	The proton energy variation with angle for protons entering the proton
	telescope is wider, but still only 5 MeV for 26 MeV protons, which allowed the
	protons to be easily identified in the E-$\Delta$E telescope. 
	
\begin{figure}
		\includegraphics[width=\linewidth]{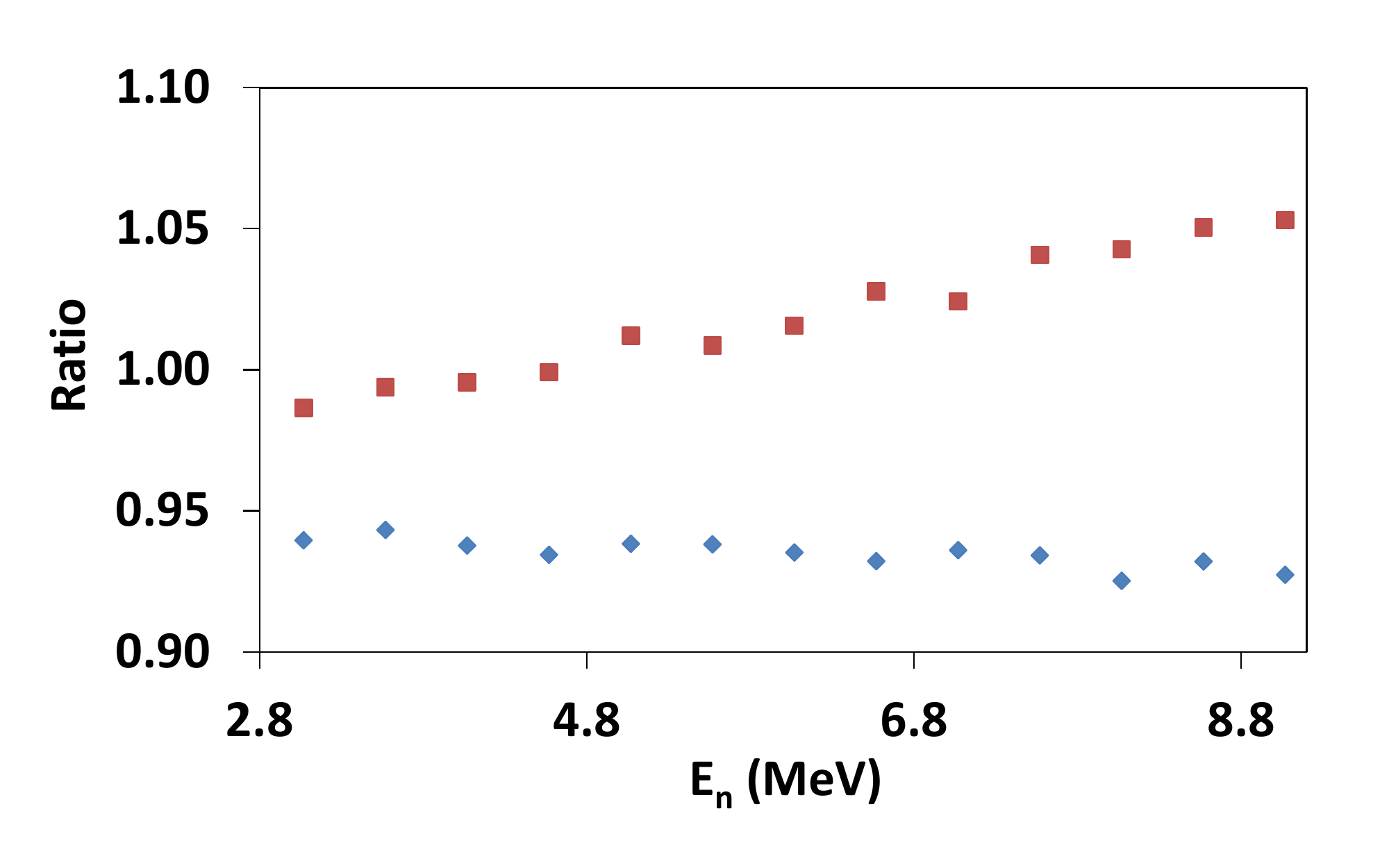}% Here is how to import EPS art
		\caption{\label{fig:ratio} The ratio of $	\left( {N_p} / {N_n} \right) $
			determined by integration using Eq.~(\ref{eq:p_telescope}) and
			Eq.~(\ref{eq:n_polytarget}) to the value determined using the simple approximate
			approach of Eq.~(\ref{eq:simple_ratio_CH2}) and Eq.~(\ref{eq:simple_ratio_G}) for the graphite(red squares) and the
			polyethylene targets (blue diamonds).  }
	\end{figure}

	Fig.~\ref{fig:ratio} shows how the corrected ratio $\left( {N_p} / {N_n} \right)
	$, computed using Eq.~(\ref{eq:p_telescope}) and Eq.~(\ref{eq:n_polytarget}), 
	compares to the approximate value determined using Eq.~(\ref{eq:simple_ratio_CH2}) and (\ref{eq:simple_ratio_G}). 
	The net effect of all these corrections was always less than about 7\%.  
	
	\subsection{Absolute Full-peak Efficiencies for Singles and Coincidence}
	
	Because the activated targets were sandwiched between the NaI detectors in order
	to maximize the count rate, the efficiency was very sensitive to geometry.  A
	Monte Carlo simulation was created using the Geant4 toolkit
	\cite{agostinelli2003,allison2006} to model both the graphite and polyethylene
	target geometries, including other decay modes, surrounding materials, and
	Compton scattering. In order to validate this code, it was used to predict
	efficiencies for other geometries which were then tested experimentally.  These
	experiments will be described in more detail in a later paper.
	
	In the first set of tests, an associated particle technique was used to allow
	the absolute singles and coincidence full-peak efficiencies to be determined
	using an uncalibrated $^{22}$Na positron source.  In this technique the
	positrons were stopped in a plastic scintillator, signaling that a pair of 511
	keV gamma rays were released.  The fraction of plastic scintillator events for
	which the NaI detector also detected a 511 keV gamma ray in coincidence is the
	absolute efficiency.  A third detector was used to correct for summing with the
	1.274 MeV gamma rays from the $^{22}$Na.  The geometry of the plastic
	scintillator and $^{22}$Na source were simulated, and compared with predicted
	singles and coincidence absolute full-peak efficiencies as a function of
	source-to-detector distance and radial source position on the face of the
	detector.  The code predictions agreed with the measurements to within
	approximately 4.7\% RMS percent difference for coincidence, and 8.6\% for
	singles.
	
	The second set of tests used an approximately 3.7 kBq NIST calibrated
	$^{68}$Ge source, with activity measured to ±1.7\%.  For the initial test, the
	source was sandwiched between two copper disks to stop the positrons, nearly
	simulating a  point source. Again the geometry was simulated and predicted
	singles and coincidence absolute full-peak efficiencies were compared with
	measurements as a function of source-to-detector distance and radial source
	position on the face of the detector, agreeing to about 5.3\% RMS percent
	difference for coincidence, and 6.6\% for singles.
	
	Finally, the $^{68}$Ge source was placed between graphite disks of the same
	diameter and approximate thickness as in the experiment, and the full-peak
	efficiencies were measured as a function of radial source position.  In this
	experiment, which is the closest to the actual measurement geometry, the
	predicted efficiencies agreed with the measurement to 6.5\% RMS percent
	difference for coincidence, and 1.4\% for singles.
	
	The GEANT simulation code was used to study the effect of misaligning at axes of
	the targets and NaI detectors.  In the worst case, which was for coincidence
	measurements using the graphite target, an offset of 1 mm of the target resulted
	a 3.1\% change in the full-peak efficiency.   
	
	Based on these results, a systematic uncertainty of 5.5\% was assigned for the
	predicted absolute full-peak efficiencies for the targets used in this
	experiment, which are listed in  Table~\ref{tab:efficiencies}.
	
	\FPeval{\GC}{0.0494}%
	\FPeval{\GCU}{round(\GC*0.055,4)}
	\FPeval{\Go}{0.0981}%
	\FPeval{\GoU}{round(\Go*0.055,4)}
	\FPeval{\Gt}{0.1049}%
	\FPeval{\GtU}{round(\Gt*0.055,4)}
	\FPeval{\PC}{0.1568}%
	\FPeval{\PCU}{round(\PC*0.05,4)}
	\FPeval{\Po}{0.1655}%
	\FPeval{\PoU}{round(\Po*0.055,4)}
	\FPeval{\Pt}{0.1656}%
	\FPeval{\PtU}{round(\Pt*0.055,4)}
	
	\begin{table}
		\caption{\label{tab:efficiencies}The absolute full-peak efficiencies used in
			the cross section calculation. }
		\begin{ruledtabular}
			\begin{tabular}{lll}
				Target&Configuration&Efficiency\\ 	\hline
				
				%	Graphite&Coincidence&$0.0487 \pm 0.00487$\\
				Graphite&Coincidence& \FPprint{\GC} $\pm$ \FPprint{\GCU} \\
				&Detector 1& \FPprint{\Go} $\pm$ \FPprint{\GoU} \\
				&Detector 2& \FPprint{\Gt} $\pm$ \FPprint{\GtU} \\
				Polyethylene&Coincidence& \FPprint{\PC} $\pm$ \FPprint{\PCU} \\
				&Detector 1& \FPprint{\Po} $\pm$ \FPprint{\PoU} \\
				&Detector 2& \FPprint{\Pt} $\pm$ \FPprint{\PtU} \\
			\end{tabular}
		\end{ruledtabular}
	\end{table}
	
	Given the large size of the graphite target used in this experiment, 
	the Monte Carlo simulation code MCNP5 \cite{mcnpteam2003} was used to reproduce
	the distribution of $^{11}$C nuclei in the graphite disk.  In this simulation,
	monoenergetic neutrons were emitted uniformly and isotropically from a circular
	region on the flat tritium target, with a radius equal to that of the last
	upstream collimator on the deuteron beam.  The interactions of these neutrons
	with the graphite disk were simulated using the standard 6000.60c cross section
	library, and resulting number of $^{11}$C nuclei were mesh tallied as a function
	of position within the graphite.  The simulation showed, relatively
	independently of the neutron energy, that the number of $^{11}$C was reduced to
	about 75\% at the downstream face relative to the upstream face of the graphite,
	and smoothly fell by approximately 5\% from the center to the edge.  Simulating
	this $^{11}$C distribution in the Geant4 code slightly increased the singles
	absolute full-peak efficiency for the detector with more $^{11}$C near it and
	reduced the other side; for coincidences the efficiency was increased by nearly
	6\%.  
	
	\subsection{Measurement Uncertainty}
	
	The  uncertainty in the cross section from  Eq.~(\ref{eq:cross_section}) was calculated using the normal propagation of uncertainty rules to get
		\begin{eqnarray}
	 \left( \delta \sigma \right)^2 & = & \sigma^2 \left\{   
	 \left( \frac{\delta N_0}{N_0} \right)^2  +
	 \left( \frac{\delta \epsilon}{\epsilon} \right)^2 + 
	 \left( \frac{\delta T_C}{T_C} \right)^2 \right. \nonumber\\
	 & & \left. + \left( \frac{\delta N_n}{N_n} \right)^2 +
	 \left( \frac{1}{\lambda} - \frac{t e^{-\lambda t}} {1-e^{-\lambda t}}  \right)^2 	 \left( \delta \lambda \right)^2    
	 \right\}
	 	\label{eq:uncert}.
	\end{eqnarray}
    where the uncertainty in the activation time $t$ is so small it has been neglected.  The range of uncertainty values for each term in Eq.~(\ref{eq:uncert}) is shown in Table~\ref{tab:uncertainties}, and the results are included in the uncertainties given in
	Table~\ref{tab:cross_sections} and Figs.~\ref{fig:cross_sections} and
	\ref{fig:ind_cross_sections}.  The estimates for statistical uncertainty in the
	number of $^{11}$C decays result from the error matrix calculated by TMinuit in
	fitting Eq.~(\ref{eq:number_of_11C}).  The range of values in
	Table~\ref{tab:uncertainties} result from the large change in the number of
	$^{11}$C produced in each target, since the  $^{12}$C(n, 2n)$^{11}$C cross
	section rises rapidly near threshold.  The uncertainty in the number of
	neutrons, determined from $\left( {N_n}/ {N_p} \right) N_p$, depends 
	on the statistical uncertainty in $N_p$ the number of protons detected (about 1.1\%), the
	uncertainty in the $^1$H(n,p)n elastic scattering cross section $\sigma_{np}$ \cite{Stoks1993,Stoks1993pion} (about  0.7\%), the uncertainty in $T_H$, the polyethylene $^1$H number density (about 3.5\%), and the solid angle uncertainties for the polyethylene and graphite targets, and the proton telescope (about 2.8\%, 1.4\%, and 2.9\% respectively). The effect of misaligning the axes of the polyethelene and graphite targets was simulated and found to result in approximately an additional  4.1\% uncertainty in the number of neutrons striking the graphite target. Since the neutrons striking the polyethylene target determines the number of protons, this systematic effect does not affect the polyethylene uncertainty.    The uncertainties in the  $^{12}$C and $^1$H areal number densities ($T_C$ and $T_H$ respectively) for each  of the targets was  estimated based on careful measurements of the target dimensions and mass.
	
	The uncertainty in the corrected incident neutron energy was assumed
	due to the uncertainty in the thickness of the tritium target, which was about
	$\pm 0.7\%$.  The neutron energy uncertainty was propagated into the  \mbox{$^{12}$C(n,
	2n)$^{11}$C} cross section uncertainty   using the slope of a polynomial fit to
	the nominal  \mbox{$^{12}$C(n, 2n)$^{11}$C} cross sections as a function of energy.  The
	resulting cross section uncertainties, which ranged from 0.31 mb to 0.85 mb, were
	added in quadrature to the other uncertainties described above.  
	
	\begin{table}
		\caption{\label{tab:uncertainties}Contribution to the cross section uncertainty
			from each term in Eq.~(\ref{eq:uncert}) corresponding to the factors in Eq.~(\ref{eq:cross_section}). For the statistical
			uncertainties in the number of $^{11}\text{C}$ and incident neutrons, the ranges
			given are for energies above about 21.5 MeV, where the cross section is large
			enough for this to be meaningful. }
		\begin{ruledtabular}
			\begin{tabular}{ll}
				&Percent\\ 	
				Source&Uncertainty\\ 	\hline \\
				Uncertainty in $N_{0}$&\\
				\hspace{1em}Graphite target& \\
				\hspace{2em}$^{11}$C decays counted in coincidence&0.3-0.9\% \\
				\hspace{2em}$^{11}$C decays counted in NaI1 detector&0.2-1.0\% \\
				\hspace{2em}$^{11}$C decays counted in NaI2 detector&0.2-1.0\% \\
				\hspace{1em}Polyethylene target& \\
				\hspace{2em}$^{11}$C decays counted in coincidence&0.7-2.3\% \\
				\hspace{2em}$^{11}$C decays counted in NaI1 detector&1.0-9.5\% \\
				\hspace{2em}$^{11}$C decays counted in NaI2 detector&1.0-5.5\% \\			
				\\ 	
				Uncertainty in NaI detector efficiency $\epsilon$ &5.5\% \\	
				\\
				Uncertainty in $^{12}$C areal number density $T_C$ & \\
				\hspace{2em}Graphite target&1.9\% \\
				\hspace{2em}Polyethylene target&3.8\% \\
				\\
				Uncertainty in incident neutrons $N_n$  & \\ 
				\hspace{2em}Graphite target&6.9\% \\
				\hspace{2em}Polyethylene target&4.6\% \\
				\\
				Uncertainty in $\left( {\lambda}/ {1-e^{-\lambda t}}\right)$&0.6\%\\ 
				
			\end{tabular}
		\end{ruledtabular}
	\end{table}

	\section{Discussion of Experimental Results}
	
	The total  $^{12}$C(n,2n)$^{11}$C cross sections obtained from the graphite and
	polyethylene targets, for coincidence and singles counting, are displayed in
	Table~\ref{tab:cross_sections}. The agreement between cross sections determined
	using different targets, geometries, efficiencies and detectors is quite good;
	the overall RMS percent difference from the mean for energies above 22 MeV,
	where there are enough statistics to be meaningful, is 6.2\%.  The best
	individual measurement is expected to be the coincidence measurement using the
	polyethylene target, since the background rate is reduced for the coincidence
	measurement, allowing a more robust fit to the growth curve, and since the same
	target is used for  $^{1}$H(n,p) elastic scattering, eliminating systematic uncertainties
	in determining the neutron flux.
	
\begin{figure*}
		\includegraphics[width=\linewidth]{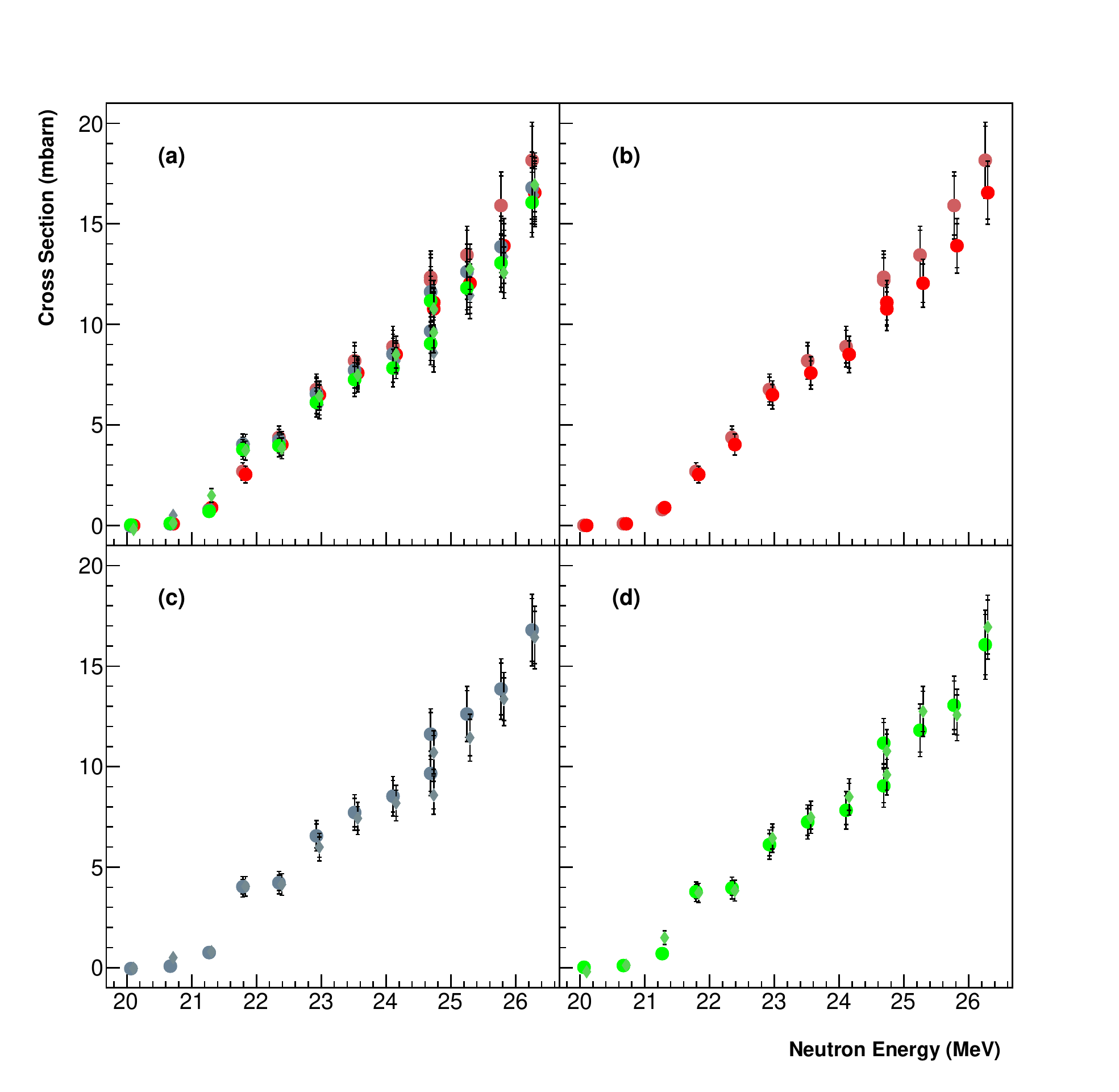}% Here is how to
		%import EPS art
		\caption{\label{fig:ind_cross_sections} Comparison of the cross sections
			measured in this experiment using the graphite (circles) and polyethylene
			(diamonds) targets for (b) both NaI detectors in coincidence (red), and (c) NaI
			detector 1 (blue) and (d) NaI detector 2 (green) in singles mode.  Figure (a)
			shows cross sections determined using both coincidence and singles information
			for both targets on the same plot.  For each cross section, the larger of the associated error bars indicates the increase in overall uncertainty
			when the uncertainty in incident neutron energy is included.}
	\end{figure*}

	Fig.~\ref{fig:ind_cross_sections} plots the cross sections measured in this
	experiment for different combinations of targets and detectors.  
	Fig.~\ref{fig:ind_cross_sections}a shows the overall agreement between all of
	the measurements made at each neutron energy. The agreement is good, but with
	the graphite coincidence measurement being systematically high.  
	Figs.~\ref{fig:ind_cross_sections}b, c and d  compare  just the polyethylene and
	graphite cross sections for coincidence and singles measurements using each
	detector.  Comparing measurements with different targets made using the same
	detector(s) shows agreement within experimental uncertainty, which is a fairly
	stringent test since the geometry and target material is not the same, supporting
	the calculated detector efficiency and geometry correction.   The
	measurement at 24.7 MeV was repeated, the first measurement made near the middle
	of the experiment and the second at the end. In this case, the coincidence measurements at the middle and end of the experiment agree when comparing graphite to graphite and polyethylene to polyethylene, but the graphite result is higher than the polyethylene for both measurements, slightly outside of the error bars. 
	
	It is clear from Fig.~\ref{fig:cross_sections} that the cross sections from the
	present experiment fall along high side of the upper curve previously set by the
	measurements of Anders et al. \cite{anders1981} and Welch et al.
	\cite{welch1981} and calculations of Dimbylow \cite{dimbylow1980}.  The
	predictions of  Dimbylow \cite{dimbylow1980} are within our error bars, as are,
	for the most part, the measurements from Anders et al  \cite{anders1981} and Welch et al. \cite{welch1981}.

 \begin{table*}
	\caption{\label{tab:cross_sections}Total  $^{12}$C(n, 2n)$^{11}$C cross sections obtained from the graphite and polyethylene targets, for coincidence and singles counting.  Also shown is the weighted mean for each energy.  The uncertainties quoted are discussed in the text.}
	\begin{ruledtabular}
		\begin{tabular}{c|cccc|ccccc}
			Deuteron&\multicolumn{4}{c|}{Graphite}&\multicolumn{4}{c}{Polyethylene} \\ 
			Energy&Energy&Coincidence&Detector 1&Detector 2&Energy&Coincidence&Detector 1&Detector 2 \\ 
			(MeV)&(MeV)&(mb)&(mb)&(mb)& (MeV) &(mb)&(mb)&(mb)\\ \hline
			3.57 & 20.06 & 0.00 $\pm$ 0.00 &-0.05 $\pm$ 0.01 &0.02 $\pm$ 0.01& 20.10  & 0.00 $\pm$ 0.01 &-0.03 $\pm$ 0.13 &-0.20 $\pm$ 0.01  \\ 
			4.07 & 20.67 & 0.08 $\pm$ 0.01 &0.07 $\pm$ 0.01 &0.10 $\pm$ 0.01& 20.71  &0.08 $\pm$ 0.01 &0.50 $\pm$ 0.11 &0.12 $\pm$ 0.01  \\ 
			4.57 & 21.27 & 0.78 $\pm$ 0.32 &0.75 $\pm$ 0.31 &0.70 $\pm$ 0.31& 21.31  &0.88 $\pm$ 0.32 &0.79 $\pm$ 0.33 &1.49 $\pm$ 0.32  \\ 
			5.07 & 21.79 & 2.69 $\pm$ 0.43 &4.03 $\pm$ 0.51 &3.78 $\pm$ 0.49& 21.83  &2.53 $\pm$ 0.41 &4.04 $\pm$ 0.49 &3.72 $\pm$ 0.41  \\ 
			5.57 & 22.35 & 4.37 $\pm$ 0.57 &4.23 $\pm$ 0.56 &3.96 $\pm$ 0.54& 22.39  &4.02 $\pm$ 0.52 &4.14 $\pm$ 0.53 &3.83 $\pm$ 0.52  \\ 
			6.07 & 22.92 & 6.76 $\pm$ 0.77 &6.56 $\pm$ 0.76 &6.12 $\pm$ 0.73& 22.97  &6.49 $\pm$ 0.70 &6.00 $\pm$ 0.69 &6.44 $\pm$ 0.70  \\ 
			6.57 & 23.51 & 8.19 $\pm$ 0.92 &7.72 $\pm$ 0.88 &7.25 $\pm$ 0.85& 23.56  &7.58 $\pm$ 0.80 &7.43 $\pm$ 0.80 &7.48 $\pm$ 0.80  \\ 
			7.07 & 24.11 & 8.89 $\pm$ 1.01 &8.53 $\pm$ 0.98 &7.83 $\pm$ 0.93& 24.15  &8.50 $\pm$ 0.90 &8.18 $\pm$ 0.88 &8.49 $\pm$ 0.90  \\ 
			7.57 & 24.69 & 12.34 $\pm$ 1.31 &11.62 $\pm$ 1.26 &11.17 $\pm$ 1.22& 24.74  &11.09 $\pm$ 1.10 &10.70 $\pm$ 1.08 &10.77 $\pm$ 1.10  \\ 
			7.57 & 24.69 & 12.19 $\pm$ 1.30 &9.66 $\pm$ 1.11 &9.04 $\pm$ 1.06& 24.74  &10.77 $\pm$ 1.08 &8.58 $\pm$ 0.95 &9.59 $\pm$ 1.08  \\ 
			8.07 & 25.25 & 13.45 $\pm$ 1.44 &12.62 $\pm$ 1.37 &11.81 $\pm$ 1.31& 25.29  &12.04 $\pm$ 1.20 &11.44 $\pm$ 1.16 &12.75 $\pm$ 1.20  \\ 
			8.57 & 25.77 & 15.91 $\pm$ 1.67 &13.86 $\pm$ 1.51 &13.05 $\pm$ 1.44& 25.82  &13.91 $\pm$ 1.36 &13.36 $\pm$ 1.32 &12.57 $\pm$ 1.36  \\ 
			9.07 & 26.25 & 18.16 $\pm$ 1.89 &16.80 $\pm$ 1.78 &16.07 $\pm$ 1.72& 26.29  &16.55 $\pm$ 1.56 &16.43 $\pm$ 1.56 &16.94 $\pm$ 1.56  \\ 
		\end{tabular}
	\end{ruledtabular}
\end{table*}

	\section{Conclusion}
	The cross sections for  $^{12}$C(n, 2n)$^{11}$C have been measured using an activation
	technique from threshold to 26.3 MeV with an uncertainty of approximately 9-12\% for the higher energies. Previous measurements disagree, tending to fall into  upper and lower
	bands.  The results of the present experiment agree with the upper band.

	  Accurate cross sections may allow the  $^{12}$C(n, 2n) reaction to be used as neutron diagnostic for ICF. During an ICF implosion, primary and tertiary DT fusion reaction neutrons are produced. Since the ion plasma temperatures in these thermonuclear implosions are typically in the keV range, the primary DT neutron energy is Q value driven. The energy distribution of primary neutrons is typically peaked around 14.1 MeV and has a small thermodynamic width which is broadened further via straggling by the compressed fuel. These neutrons are copious and exceed the production of tertiary neutrons by six to seven orders of magnitude. Conversely, neutrons which are produced by DT fusion reactions caused by up-scattered MeV DT fuel generate tertiary neutrons with energies in the 10 to 32 MeV range. Since the  $^{12}$C(n, 2n) reaction is insensitive to energies below 20 MeV,  the primary 14.1 MeV DT and down scattered primary neutrons cannot react with the carbon. Only tertiary neutrons in 20 to 32 MeV range react making this method useful as a possible ICF tertiary neutron diagnostic. 
	
	To use this method, ultra-pure graphite disks placed within the ICF reaction chamber become activated by tertiary neutrons via the  $^{12}$C(n, 2n)$^{11}$C reaction. The $^{11}$C in the disk subsequently decays via positron emission, and the 511 keV annihilation gamma rays  are then counted in coincidence using a detector system far   from the target chamber.  The gamma counts can then be used to obtain the tertiary neutron yield, limited by the uncertainty in the  $^{12}$C(n, 2n)$^{11}$C cross section.

	\begin{acknowledgments}
		We would like to thank Andrew Evans, Keith Mann, Tyler Reynolds, Ian Love,
		August Gula, Laurel Vincett, Lee Gabler, Michael Krieger, Mollie Bienstock,
		Collin Stillman, Drew Ellison, and Holly Desmitt for assistance performing the
		experiment, and Shamim Akhtar, Don Carter, Sushil Dhakal, Devon Jacobs, John
		O'Donnell, and Andrea Richard for assistance at Ohio University. This research is
		funded in part by the University of Rochester Laboratory for Laser Energetics
		through a grant from the Department of Energy and by the U.S. Department of
		Energy, under Grant Nos. DE-FG02-88ER40387, DE-NA0001837, and DE-NA0002905.
		
		Certain commercial equipment, instruments, or materials are identified in this paper to foster understanding. Such identification does not imply recommendation or endorsement by the National Institute of Standards and Technology, nor does it imply that the materials or equipment identified are necessarily the best available for the purpose.
	\end{acknowledgments}
	
	%\newpage %Just because of unusual number of tables stacked at end
	\bibliography{./references/12C(n2n)}% Produces the bibliography via BibTeX.

%merlin.mbs apsrev4-1.bst 2010-07-25 4.21a (PWD, AO, DPC) hacked
%Control: key (0)
%Control: author (8) initials jnrlst
%Control: editor formatted (1) identically to author
%Control: production of article title (-1) disabled
%Control: page (0) single
%Control: year (1) truncated
%Control: production of eprint (0) enabled
\begin{thebibliography}{21}%
\makeatletter
\providecommand \@ifxundefined [1]{%
 \@ifx{#1\undefined}
}%
\providecommand \@ifnum [1]{%
 \ifnum #1\expandafter \@firstoftwo
 \else \expandafter \@secondoftwo
 \fi
}%
\providecommand \@ifx [1]{%
 \ifx #1\expandafter \@firstoftwo
 \else \expandafter \@secondoftwo
 \fi
}%
\providecommand \natexlab [1]{#1}%
\providecommand \enquote  [1]{``#1''}%
\providecommand \bibnamefont  [1]{#1}%
\providecommand \bibfnamefont [1]{#1}%
\providecommand \citenamefont [1]{#1}%
\providecommand \href@noop [0]{\@secondoftwo}%
\providecommand \href [0]{\begingroup \@sanitize@url \@href}%
\providecommand \@href[1]{\@@startlink{#1}\@@href}%
\providecommand \@@href[1]{\endgroup#1\@@endlink}%
\providecommand \@sanitize@url [0]{\catcode `\\12\catcode `\$12\catcode
  `\&12\catcode `\#12\catcode `\^12\catcode `\_12\catcode `\%12\relax}%
\providecommand \@@startlink[1]{}%
\providecommand \@@endlink[0]{}%
\providecommand \url  [0]{\begingroup\@sanitize@url \@url }%
\providecommand \@url [1]{\endgroup\@href {#1}{\urlprefix }}%
\providecommand \urlprefix  [0]{URL }%
\providecommand \Eprint [0]{\href }%
\providecommand \doibase [0]{http://dx.doi.org/}%
\providecommand \selectlanguage [0]{\@gobble}%
\providecommand \bibinfo  [0]{\@secondoftwo}%
\providecommand \bibfield  [0]{\@secondoftwo}%
\providecommand \translation [1]{[#1]}%
\providecommand \BibitemOpen [0]{}%
\providecommand \bibitemStop [0]{}%
\providecommand \bibitemNoStop [0]{.\EOS\space}%
\providecommand \EOS [0]{\spacefactor3000\relax}%
\providecommand \BibitemShut  [1]{\csname bibitem#1\endcsname}%
\let\auto@bib@innerbib\@empty
%</preamble>
\bibitem [{\citenamefont {Welch}\ \emph {et~al.}(1988)\citenamefont {Welch},
  \citenamefont {Kislev},\ and\ \citenamefont {Miley}}]{welch1988}%
  \BibitemOpen
  \bibfield  {author} {\bibinfo {author} {\bibfnamefont {D.}~\bibnamefont
  {Welch}}, \bibinfo {author} {\bibfnamefont {H.}~\bibnamefont {Kislev}}, \
  and\ \bibinfo {author} {\bibfnamefont {G.}~\bibnamefont {Miley}},\
  }\href@noop {} {\bibfield  {journal} {\bibinfo  {journal} {Rev Sci Instrum}\
  }\textbf {\bibinfo {volume} {59}},\ \bibinfo {pages} {610} (\bibinfo {year}
  {1988})}\BibitemShut {NoStop}%
\bibitem [{\citenamefont {Glebov}\ \emph {et~al.}(2003)\citenamefont {Glebov},
  \citenamefont {Stoeckl}, \citenamefont {Sangster}, \citenamefont
  {Meyerhofer}, \citenamefont {Radha}, \citenamefont {Padalino}, \citenamefont
  {Baumgart}, \citenamefont {Colburn},\ and\ \citenamefont
  {Fuschino}}]{glebov2003}%
  \BibitemOpen
  \bibfield  {author} {\bibinfo {author} {\bibfnamefont {V.~Y.}\ \bibnamefont
  {Glebov}}, \bibinfo {author} {\bibfnamefont {C.}~\bibnamefont {Stoeckl}},
  \bibinfo {author} {\bibfnamefont {T.}~\bibnamefont {Sangster}}, \bibinfo
  {author} {\bibfnamefont {D.}~\bibnamefont {Meyerhofer}}, \bibinfo {author}
  {\bibfnamefont {P.}~\bibnamefont {Radha}}, \bibinfo {author} {\bibfnamefont
  {S.}~\bibnamefont {Padalino}}, \bibinfo {author} {\bibfnamefont
  {L.}~\bibnamefont {Baumgart}}, \bibinfo {author} {\bibfnamefont
  {R.}~\bibnamefont {Colburn}}, \ and\ \bibinfo {author} {\bibfnamefont
  {J.}~\bibnamefont {Fuschino}},\ }\href@noop {} {\bibfield  {journal}
  {\bibinfo  {journal} {Rev Sci Instrum}\ }\textbf {\bibinfo {volume} {74}},\
  \bibinfo {pages} {1717} (\bibinfo {year} {2003})}\BibitemShut {NoStop}%
\bibitem [{\citenamefont {Galbiati}\ \emph {et~al.}(2005)\citenamefont
  {Galbiati}, \citenamefont {Pocar}, \citenamefont {Franco}, \citenamefont
  {Ianni}, \citenamefont {Cadonati},\ and\ \citenamefont
  {Sch{\"o}nert}}]{galbiati2005}%
  \BibitemOpen
  \bibfield  {author} {\bibinfo {author} {\bibfnamefont {C.}~\bibnamefont
  {Galbiati}}, \bibinfo {author} {\bibfnamefont {A.}~\bibnamefont {Pocar}},
  \bibinfo {author} {\bibfnamefont {D.}~\bibnamefont {Franco}}, \bibinfo
  {author} {\bibfnamefont {A.}~\bibnamefont {Ianni}}, \bibinfo {author}
  {\bibfnamefont {L.}~\bibnamefont {Cadonati}}, \ and\ \bibinfo {author}
  {\bibfnamefont {S.}~\bibnamefont {Sch{\"o}nert}},\ }\href@noop {} {\bibfield
  {journal} {\bibinfo  {journal} {Phys. Rev. C}\ }\textbf {\bibinfo {volume}
  {71}},\ \bibinfo {pages} {055805} (\bibinfo {year} {2005})}\BibitemShut
  {NoStop}%
\bibitem [{\citenamefont {Audi}\ \emph {et~al.}(2012)\citenamefont {Audi},
  \citenamefont {Kondev}, \citenamefont {Wang}, \citenamefont {Pfeiffer},
  \citenamefont {Sun}, \citenamefont {Blachot},\ and\ \citenamefont
  {MacCormick}}]{NuBase}%
  \BibitemOpen
  \bibfield  {author} {\bibinfo {author} {\bibfnamefont {G.}~\bibnamefont
  {Audi}}, \bibinfo {author} {\bibfnamefont {F.}~\bibnamefont {Kondev}},
  \bibinfo {author} {\bibfnamefont {M.}~\bibnamefont {Wang}}, \bibinfo {author}
  {\bibfnamefont {B.}~\bibnamefont {Pfeiffer}}, \bibinfo {author}
  {\bibfnamefont {X.}~\bibnamefont {Sun}}, \bibinfo {author} {\bibfnamefont
  {J.}~\bibnamefont {Blachot}}, \ and\ \bibinfo {author} {\bibfnamefont
  {M.}~\bibnamefont {MacCormick}},\ }\href
  {http://stacks.iop.org/1674-1137/36/i=12/a=001} {\bibfield  {journal}
  {\bibinfo  {journal} {Chin. Phys. C}\ }\textbf {\bibinfo {volume} {36}},\
  \bibinfo {pages} {1157} (\bibinfo {year} {2012})}\BibitemShut {NoStop}%
\bibitem [{\citenamefont {Dimbylow}(1980)}]{dimbylow1980}%
  \BibitemOpen
  \bibfield  {author} {\bibinfo {author} {\bibfnamefont {P.}~\bibnamefont
  {Dimbylow}},\ }\href@noop {} {\bibfield  {journal} {\bibinfo  {journal}
  {Phys. Med. Biol.}\ }\textbf {\bibinfo {volume} {25}},\ \bibinfo {pages}
  {637} (\bibinfo {year} {1980})}\BibitemShut {NoStop}%
\bibitem [{\citenamefont {Anders}\ \emph {et~al.}(1981)\citenamefont {Anders},
  \citenamefont {Herges},\ and\ \citenamefont {Scobel}}]{anders1981}%
  \BibitemOpen
  \bibfield  {author} {\bibinfo {author} {\bibfnamefont {B.}~\bibnamefont
  {Anders}}, \bibinfo {author} {\bibfnamefont {P.}~\bibnamefont {Herges}}, \
  and\ \bibinfo {author} {\bibfnamefont {W.}~\bibnamefont {Scobel}},\
  }\href@noop {} {\bibfield  {journal} {\bibinfo  {journal} {Z. Phys. A}\
  }\textbf {\bibinfo {volume} {301}},\ \bibinfo {pages} {353} (\bibinfo {year}
  {1981})}\BibitemShut {NoStop}%
\bibitem [{\citenamefont {Welch}\ \emph {et~al.}(1981)\citenamefont {Welch},
  \citenamefont {Johnson}, \citenamefont {Randers-Pehrson},\ and\ \citenamefont
  {Rapaport}}]{welch1981}%
  \BibitemOpen
  \bibfield  {author} {\bibinfo {author} {\bibfnamefont {P.}~\bibnamefont
  {Welch}}, \bibinfo {author} {\bibfnamefont {J.}~\bibnamefont {Johnson}},
  \bibinfo {author} {\bibfnamefont {G.}~\bibnamefont {Randers-Pehrson}}, \ and\
  \bibinfo {author} {\bibfnamefont {J.}~\bibnamefont {Rapaport}},\ }\href@noop
  {} {\bibfield  {journal} {\bibinfo  {journal} {Bull. Am. Phys. Soc}\ }\textbf
  {\bibinfo {volume} {26}},\ \bibinfo {pages} {708} (\bibinfo {year}
  {1981})}\BibitemShut {NoStop}%
\bibitem [{\citenamefont {Brolley~Jr}\ \emph {et~al.}(1952)\citenamefont
  {Brolley~Jr}, \citenamefont {Fowler},\ and\ \citenamefont
  {Schlacks}}]{brolley1952}%
  \BibitemOpen
  \bibfield  {author} {\bibinfo {author} {\bibfnamefont {J.}~\bibnamefont
  {Brolley~Jr}}, \bibinfo {author} {\bibfnamefont {J.}~\bibnamefont {Fowler}},
  \ and\ \bibinfo {author} {\bibfnamefont {L.}~\bibnamefont {Schlacks}},\
  }\href@noop {} {\bibfield  {journal} {\bibinfo  {journal} {Phys. Rev.}\
  }\textbf {\bibinfo {volume} {88}},\ \bibinfo {pages} {618} (\bibinfo {year}
  {1952})}\BibitemShut {NoStop}%
\bibitem [{\citenamefont {Brill}\ \emph {et~al.}(1961)\citenamefont {Brill},
  \citenamefont {Vlasov}, \citenamefont {Kalinin},\ and\ \citenamefont
  {Sokolov}}]{brill1961}%
  \BibitemOpen
  \bibfield  {author} {\bibinfo {author} {\bibfnamefont {O.~D.}\ \bibnamefont
  {Brill}}, \bibinfo {author} {\bibfnamefont {N.~A.}\ \bibnamefont {Vlasov}},
  \bibinfo {author} {\bibfnamefont {S.~P.}\ \bibnamefont {Kalinin}}, \ and\
  \bibinfo {author} {\bibfnamefont {L.~S.}\ \bibnamefont {Sokolov}},\
  }\href@noop {} {\bibfield  {journal} {\bibinfo  {journal} {Dok. Akad. Nauk
  SSSR}\ }\textbf {\bibinfo {volume} {136}},\ \bibinfo {pages} {55} (\bibinfo
  {year} {1961})},\ \bibinfo {note} {[Sov. Phys. Dokl. 6, 24-26
  (1961)].}\BibitemShut {Stop}%
\bibitem [{\citenamefont {Soewarsono}\ \emph {et~al.}(1992)\citenamefont
  {Soewarsono}, \citenamefont {Uwamino},\ and\ \citenamefont
  {Nakamura}}]{soewarsono1992}%
  \BibitemOpen
  \bibfield  {author} {\bibinfo {author} {\bibfnamefont {T.~S.}\ \bibnamefont
  {Soewarsono}}, \bibinfo {author} {\bibfnamefont {Y.}~\bibnamefont {Uwamino}},
  \ and\ \bibinfo {author} {\bibfnamefont {T.}~\bibnamefont {Nakamura}},\ }in\
  \href@noop {} {\emph {\bibinfo {booktitle} {Proceedings of the 1991 Symposium
  on Nuclear Data}}},\ \bibinfo {series and number} {\bibinfo {number} {JAERI-M
  92-027}},\ \bibinfo {editor} {edited by\ \bibinfo {editor} {\bibfnamefont
  {M.}~\bibnamefont {Baba}}\ and\ \bibinfo {editor} {\bibfnamefont
  {T.}~\bibnamefont {Nakagawa}}}\ (\bibinfo  {publisher} {Tokyo},\ \bibinfo
  {address} {Tokai, Ibaraki-ken. Japan},\ \bibinfo {year} {1992})\ pp.\
  \bibinfo {pages} {354--363}\BibitemShut {NoStop}%
\bibitem [{\citenamefont {Uno}\ \emph {et~al.}(1996)\citenamefont {Uno},
  \citenamefont {Uwamino}, \citenamefont {Soewarsono},\ and\ \citenamefont
  {Nakamura}}]{uno1996}%
  \BibitemOpen
  \bibfield  {author} {\bibinfo {author} {\bibfnamefont {Y.}~\bibnamefont
  {Uno}}, \bibinfo {author} {\bibfnamefont {Y.}~\bibnamefont {Uwamino}},
  \bibinfo {author} {\bibfnamefont {T.~S.}\ \bibnamefont {Soewarsono}}, \ and\
  \bibinfo {author} {\bibfnamefont {T.}~\bibnamefont {Nakamura}},\ }\href@noop
  {} {\bibfield  {journal} {\bibinfo  {journal} {Nucl Sci Eng}\ }\textbf
  {\bibinfo {volume} {122}},\ \bibinfo {pages} {247} (\bibinfo {year}
  {1996})}\BibitemShut {NoStop}%
\bibitem [{\citenamefont {Kelley}\ \emph {et~al.}(2012)\citenamefont {Kelley},
  \citenamefont {Kwan}, \citenamefont {Purcell}, \citenamefont {Sheu},\ and\
  \citenamefont {Weller}}]{kelley2012}%
  \BibitemOpen
  \bibfield  {author} {\bibinfo {author} {\bibfnamefont {J.~H.}\ \bibnamefont
  {Kelley}}, \bibinfo {author} {\bibfnamefont {E.}~\bibnamefont {Kwan}},
  \bibinfo {author} {\bibfnamefont {J.}~\bibnamefont {Purcell}}, \bibinfo
  {author} {\bibfnamefont {C.}~\bibnamefont {Sheu}}, \ and\ \bibinfo {author}
  {\bibfnamefont {H.}~\bibnamefont {Weller}},\ }\href@noop {} {\bibfield
  {journal} {\bibinfo  {journal} {Nucl. Phys. A}\ }\textbf {\bibinfo {volume}
  {880}},\ \bibinfo {pages} {88} (\bibinfo {year} {2012})}\BibitemShut
  {NoStop}%
\bibitem [{\citenamefont {Drosg}(2000)}]{drosg2000}%
  \BibitemOpen
  \bibfield  {author} {\bibinfo {author} {\bibfnamefont {M.}~\bibnamefont
  {Drosg}},\ }\href@noop {} {\emph {\bibinfo {title} {DROSG-2000: Neutron
  source reactions.}}},\ \bibinfo {type} {Tech. Rep.}\ \bibinfo {number}
  {IAEA-NDS-87 Rev. 5}\ (\bibinfo  {institution} {International Atomic Energy
  Agency, Nuclear Data Section, Vienna (Austria)},\ \bibinfo {year}
  {2000})\BibitemShut {NoStop}%
\bibitem [{\citenamefont {Stoks}\ \emph
  {et~al.}(1993{\natexlab{a}})\citenamefont {Stoks}, \citenamefont {Klomp},
  \citenamefont {Rentmeester},\ and\ \citenamefont {De~Swart}}]{Stoks1993}%
  \BibitemOpen
  \bibfield  {author} {\bibinfo {author} {\bibfnamefont {V.}~\bibnamefont
  {Stoks}}, \bibinfo {author} {\bibfnamefont {R.}~\bibnamefont {Klomp}},
  \bibinfo {author} {\bibfnamefont {M.}~\bibnamefont {Rentmeester}}, \ and\
  \bibinfo {author} {\bibfnamefont {J.}~\bibnamefont {De~Swart}},\ }\href@noop
  {} {\bibfield  {journal} {\bibinfo  {journal} {Phys. Rev. C}\ }\textbf
  {\bibinfo {volume} {48}},\ \bibinfo {pages} {792} (\bibinfo {year}
  {1993}{\natexlab{a}})}\BibitemShut {NoStop}%
\bibitem [{\citenamefont {Stoks}\ \emph {et~al.}(1994)\citenamefont {Stoks},
  \citenamefont {Klomp}, \citenamefont {Terheggen},\ and\ \citenamefont
  {De~Swart}}]{Stoks1994}%
  \BibitemOpen
  \bibfield  {author} {\bibinfo {author} {\bibfnamefont {V.}~\bibnamefont
  {Stoks}}, \bibinfo {author} {\bibfnamefont {R.}~\bibnamefont {Klomp}},
  \bibinfo {author} {\bibfnamefont {C.}~\bibnamefont {Terheggen}}, \ and\
  \bibinfo {author} {\bibfnamefont {J.}~\bibnamefont {De~Swart}},\ }\href@noop
  {} {\bibfield  {journal} {\bibinfo  {journal} {Phys. Rev. C}\ }\textbf
  {\bibinfo {volume} {49}},\ \bibinfo {pages} {2950} (\bibinfo {year}
  {1994})}\BibitemShut {NoStop}%
\bibitem [{\citenamefont {Brun}\ and\ \citenamefont
  {Rademakers}(1997)}]{Brun1997}%
  \BibitemOpen
  \bibfield  {author} {\bibinfo {author} {\bibfnamefont {R.}~\bibnamefont
  {Brun}}\ and\ \bibinfo {author} {\bibfnamefont {F.}~\bibnamefont
  {Rademakers}},\ }\href@noop {} {\bibfield  {journal} {\bibinfo  {journal}
  {Nucl. Instrum. Methods Phys. Res. A}\ }\textbf {\bibinfo {volume} {389}},\
  \bibinfo {pages} {81} (\bibinfo {year} {1997})}\BibitemShut {NoStop}%
\bibitem [{\citenamefont {James}\ and\ \citenamefont
  {Winkler}(1998)}]{james1998}%
  \BibitemOpen
  \bibfield  {author} {\bibinfo {author} {\bibfnamefont {F.}~\bibnamefont
  {James}}\ and\ \bibinfo {author} {\bibfnamefont {M.}~\bibnamefont
  {Winkler}},\ }\href@noop {} {\emph {\bibinfo {title} {MINUIT--Users Guide,
  Program Library D506}}},\ \bibinfo {organization} {CERN, Geneva} (\bibinfo
  {year} {1998})\BibitemShut {NoStop}%
\bibitem [{\citenamefont {Agostinelli}\ \emph {et~al.}(2003)\citenamefont
  {Agostinelli}, \citenamefont {Allison}, \citenamefont {Amako}, \citenamefont
  {Apostolakis}, \citenamefont {Araujo}, \citenamefont {Arce}, \citenamefont
  {Asai}, \citenamefont {Axen}, \citenamefont {Banerjee}, \citenamefont
  {Barrand} \emph {et~al.}}]{agostinelli2003}%
  \BibitemOpen
  \bibfield  {author} {\bibinfo {author} {\bibfnamefont {S.}~\bibnamefont
  {Agostinelli}}, \bibinfo {author} {\bibfnamefont {J.}~\bibnamefont
  {Allison}}, \bibinfo {author} {\bibfnamefont {K.~a.}\ \bibnamefont {Amako}},
  \bibinfo {author} {\bibfnamefont {J.}~\bibnamefont {Apostolakis}}, \bibinfo
  {author} {\bibfnamefont {H.}~\bibnamefont {Araujo}}, \bibinfo {author}
  {\bibfnamefont {P.}~\bibnamefont {Arce}}, \bibinfo {author} {\bibfnamefont
  {M.}~\bibnamefont {Asai}}, \bibinfo {author} {\bibfnamefont {D.}~\bibnamefont
  {Axen}}, \bibinfo {author} {\bibfnamefont {S.}~\bibnamefont {Banerjee}},
  \bibinfo {author} {\bibfnamefont {G.}~\bibnamefont {Barrand}},  \emph
  {et~al.},\ }\href@noop {} {\bibfield  {journal} {\bibinfo  {journal} {Nucl.
  Instrum. Methods Phys. Res. A}\ }\textbf {\bibinfo {volume} {506}},\ \bibinfo
  {pages} {250} (\bibinfo {year} {2003})}\BibitemShut {NoStop}%
\bibitem [{\citenamefont {Allison}\ \emph {et~al.}(2006)\citenamefont
  {Allison}, \citenamefont {Amako}, \citenamefont {Apostolakis}, \citenamefont
  {Araujo}, \citenamefont {Dubois}, \citenamefont {Asai}, \citenamefont
  {Barrand}, \citenamefont {Capra}, \citenamefont {Chauvie}, \citenamefont
  {Chytracek} \emph {et~al.}}]{allison2006}%
  \BibitemOpen
  \bibfield  {author} {\bibinfo {author} {\bibfnamefont {J.}~\bibnamefont
  {Allison}}, \bibinfo {author} {\bibfnamefont {K.}~\bibnamefont {Amako}},
  \bibinfo {author} {\bibfnamefont {J.~e.~a.}\ \bibnamefont {Apostolakis}},
  \bibinfo {author} {\bibfnamefont {H.}~\bibnamefont {Araujo}}, \bibinfo
  {author} {\bibfnamefont {P.~A.}\ \bibnamefont {Dubois}}, \bibinfo {author}
  {\bibfnamefont {M.}~\bibnamefont {Asai}}, \bibinfo {author} {\bibfnamefont
  {G.}~\bibnamefont {Barrand}}, \bibinfo {author} {\bibfnamefont
  {R.}~\bibnamefont {Capra}}, \bibinfo {author} {\bibfnamefont
  {S.}~\bibnamefont {Chauvie}}, \bibinfo {author} {\bibfnamefont
  {R.}~\bibnamefont {Chytracek}},  \emph {et~al.},\ }\href@noop {} {\bibfield
  {journal} {\bibinfo  {journal} {Nuclear Science, IEEE Transactions on}\
  }\textbf {\bibinfo {volume} {53}},\ \bibinfo {pages} {270} (\bibinfo {year}
  {2006})}\BibitemShut {NoStop}%
\bibitem [{\citenamefont {Team}(2003)}]{mcnpteam2003}%
  \BibitemOpen
  \bibfield  {author} {\bibinfo {author} {\bibfnamefont {X.}~\bibnamefont
  {Team}},\ }\href@noop {} {\bibfield  {journal} {\bibinfo  {journal} {Los
  Alamos National Lab Report: LA-UR-03-1987}\ } (\bibinfo {year}
  {2003})}\BibitemShut {NoStop}%
\bibitem [{\citenamefont {Stoks}\ \emph
  {et~al.}(1993{\natexlab{b}})\citenamefont {Stoks}, \citenamefont
  {Timmermans},\ and\ \citenamefont {De~Swart}}]{Stoks1993pion}%
  \BibitemOpen
  \bibfield  {author} {\bibinfo {author} {\bibfnamefont {V.}~\bibnamefont
  {Stoks}}, \bibinfo {author} {\bibfnamefont {R.}~\bibnamefont {Timmermans}}, \
  and\ \bibinfo {author} {\bibfnamefont {J.}~\bibnamefont {De~Swart}},\
  }\href@noop {} {\bibfield  {journal} {\bibinfo  {journal} {Phys. Rev. C}\
  }\textbf {\bibinfo {volume} {47}},\ \bibinfo {pages} {512} (\bibinfo {year}
  {1993}{\natexlab{b}})}\BibitemShut {NoStop}%
\end{thebibliography}%
	
\end{document}